\documentclass[usenatbib,fleqn,a4paper]{mn2e}
\usepackage{times}
\usepackage{txfonts}
\usepackage{graphicx}
\usepackage{setspace}
\usepackage{natbib}
\usepackage{sidecap}
\usepackage{color}
\usepackage{subfigure}
\usepackage{floatflt}
\usepackage{rotating}
\usepackage{multirow}
\usepackage{aalongtable, lscape, afterpage}

\begin{document}


\newcommand{\st}[1]{\mathrm{#1}} 
\newcommand{\pow}[2]{$\st{#1}^{#2}$}
\newcommand{\grad}{\hspace{-0.15em}\r{}}
\newcommand{\lm}{\lambda}
\newcommand{\average}[1]{\left\langle #1 \right\rangle}
\newcommand{\pp}[1]{#1^{\prime\prime}}
\newcommand{\p}[1]{#1^{\prime}}

\newcommand{\fobs}{F$_{\st {obs}}$}
\newcommand{\fintrinsic}{F$_{\st {intr}}$}
\newcommand{\fgal}{F$_{\st {gal}}$}

\newcommand{\oi}{[O{\sc i}]}
\newcommand{\cii}{[C{\sc ii}]}
\newcommand{\nii}{[N{\sc ii}]}
\newcommand{\oiii}{[O{\sc iii}]}
\newcommand{\oii}{[O{\sc ii}]}
\newcommand{\si}{[Si{\sc i}]}
\newcommand{\mm}{~$\mu$m}
\newcommand{\mminv}{\mm$^{-1}$}
\newcommand{\ha}{\st{H}\alpha}
\newcommand{\hb}{\st{H}\beta}
\newcommand{\civ}{[C{\sc iv}]}
\newcommand{\niiopt}{[N{\sc ii}]~$\lambda6583$}
\newcommand{\niia}{[N{\sc ii}]~$\lambda5199$}
\newcommand{\siia}{[S{\sc ii}]~$\lambda6717$}
\newcommand{\siib}{[S{\sc ii}]~$\lambda6731$}
\newcommand{\oiiopt}{[O{\sc ii}]~$\lambda3727$}
\newcommand{\oia}{[O{\sc i}]~$\lambda6300$}
\newcommand{\oib}{[O{\sc i}]~$\lambda6363$}
\newcommand{\oiiiopt}{[O{\sc iii}]~$\lambda5007$}
\newcommand{\ohb}{[\st{O_{~III}}]~\lambda~5007/{\st H}\beta}

\newcommand{\md}{M_{\st{d}}}
\newcommand{\td}{T_{\st{d}}}
\newcommand{\mdc}{M_{\st{d,c}}}
\newcommand{\tdc}{T_{\st{d,c}}}
\newcommand{\mdw}{M_{\st{d,w}}}
\newcommand{\tdw}{T_{\st{d,w}}}
\newcommand{\mg}{M_{\st{g}}}
\newcommand{\tcmb}{T_{\st{cmb}}}
\newcommand{\hextra}{H_{\st{extra}}}
\newcommand{\hcol}{N$_{\st H}$}
\newcommand{\pdv}{$P{\st d}V$}
\newcommand{\nitrogen}{$Z_{\odot}$(N)}
\newcommand{\mh}{M_{\st H_2}}

\newcommand{\h}{~h_{71}~}
\newcommand{\hinv}[1]{~h_{71}^{#1}~}
\newcommand{\eq}[1]{(\ref{eq-#1})}
\newcommand{\wrt}{with respect to\ }
\newcommand{\mms}{\frac{M_{\odot}}{M}}
\newcommand{\mpy}{\ms~\st{yr}^{-1}}
\newcommand{\ct}{t_{\st{cool}}}
\newcommand{\rc}{r_{\st{cool}}}
\newcommand{\lc}{L_{\st{cool}}}
\newcommand{\tvir}{T_{\st{vir}}}
\newcommand{\rvir}{R_{\st{500}}}
\newcommand{\mvir}{M_{\st{500}}}
\newcommand{\mbh}{M_{\st{BH}}}
\newcommand{\ms}{M_{\odot}}
\newcommand{\ls}{L_{\odot}}
\newcommand{\zs}{Z_{\odot}}
\newcommand{\lxb}{L_{\st {Xb}}}
\newcommand{\lfir}{L_{\st{FIR}}}
\newcommand{\lfirtot}{L_{\st{FIR,tot}}}
\newcommand{\lx}{L_{\st {X}}}
\newcommand{\lr}{L_{\st {R}}}
\newcommand{\rlum}{L_{\nu~(1.4~\textrm{GHz})}}
\newcommand{\lbcg}{L_{\st{BCG}}}
\newcommand{\lt}{L_{\st{X}}{\st -}\tvir}
\newcommand{\Dl}{D_{\st{L}}}
\newcommand{\Da}{D_{\st{A}}}
\newcommand{\mdr}{\dot{M}_{\st{classical}}}
\newcommand{\smdr}{\dot{M}_{\st{spec}}}
\newcommand{\norm}{$\eta_{\st{OSP}}$}

\newcommand{\hydra}{Hydra-A}
\newcommand{\pks}{PKS~0745-191}
\newcommand{\rxj}{RXC~J1504}
\newcommand{\zw}{ZwCl~3146}
\newcommand{\pers}{NGC~1275}
\newcommand{\cen}{NGC~4696}
\newcommand{\tosp}{\tau_{\st{o}}}
\newcommand{\tysp}{\tau_{\st{y}}}
\newcommand{\tyspold}{\tau_{\st{y,old}}}
\newcommand{\mosp}{M_{\st{o}}}
\newcommand{\mysp}{M_{\st{y}}}
\newcommand{\zosp}{Z_{\st {osp}}}
\newcommand{\zysp}{Z_{\st {ysp}}}
\newcommand{\nbursts}{N_{\textrm{\small bursts}}}
\newcommand{\lamvec}{{\theta}}
\newcommand{\Fis}{\{F_i\}}
\newcommand{\text}[1]{{\rm #1}}

\newcommand{\chandra}{\textit{Chandra}}
\newcommand{\vla}{\textit{VLA}}
\newcommand{\gmrt}{\textit{GMRT}}
\newcommand{\atca}{\textit{ATCA}}
\newcommand{\XMM}{\textit{XMM-Newton}}
\newcommand{\einstein}{\textit{Einstein}}
\newcommand{\asca}{\textit{ASCA}}
\newcommand{\rosat}{\textit{ROSAT}}
\newcommand{\herschel}{\textit{Herschel}}
\newcommand{\iras}{\textit{IRAS}}
\newcommand{\spitzer}{\textit{Spitzer}}
\newcommand{\hiflux}{\textit{HIFLUGCS}}

\newcommand{\combf}[1]{{\bf #1}}

\newcommand{\bc}{\sc bayescool}
%
\def\aj{AJ}%
\def\araa{ARA\&A}%
\def\apj{ApJ}%
\def\apjl{ApJ}%
\def\apjs{ApJS}%
\def\ao{Appl.~Opt.}%
\def\apss{Ap\&SS}%
\def\aap{A\&A}%
\def\aapr{A\&A~Rev.}%
\def\aaps{A\&AS}%
\def\azh{AZh}%
\def\baas{BAAS}%
\def\jrasc{JRASC}%
\def\memras{MmRAS}%
\def\mnras{MNRAS}%
\def\pra{Phys.~Rev.~A}%
\def\prb{Phys.~Rev.~B}%
\def\prc{Phys.~Rev.~C}%
\def\prd{Phys.~Rev.~D}%
\def\pre{Phys.~Rev.~E}%
\def\prl{Phys.~Rev.~Lett.}%
\def\pasp{PASP}%
\def\pasj{PASJ}%
\def\qjras{QJRAS}%
\def\skytel{S\&T}%
\def\solphys{Sol.~Phys.}%
\def\sovast{Soviet~Ast.}%
\def\ssr{Space~Sci.~Rev.}%
\def\zap{ZAp}%
\def\nat{Nature}%
\def\iaucirc{IAU~Circ.}%
\def\aplett{Astrophys.~Lett.}%
\def\apspr{Astrophys.~Space~Phys.~Res.}%
\def\bain{Bull.~Astron.~Inst.~Netherlands}%
\def\fcp{Fund.~Cosmic~Phys.}%
\def\gca{Geochim.~Cosmochim.~Acta}%
\def\grl{Geophys.~Res.~Lett.}%
\def\jcp{J.~Chem.~Phys.}%
\def\jgr{J.~Geophys.~Res.}%
\def\jqsrt{J.~Quant.~Spec.~Radiat.~Transf.}%
\def\memsai{Mem.~Soc.~Astron.~Italiana}%
\def\nphysa{Nucl.~Phys.~A}%
\def\physrep{Phys.~Rep.}%
\def\physscr{Phys.~Scr}%
\def\planss{Planet.~Space~Sci.}%
\def\procspie{Proc.~SPIE}%

\title [Star Formation Rate in the BCG of Phoenix] {The challenging task of determining star formation rates: \\the case of
  a massive stellar burst in the brightest cluster galaxy of
  Phoenix galaxy cluster } 
\author[Rupal Mittal]{ Rupal~Mittal$^{1,2}$\thanks{E-mail: rmittal@astro.rit.edu}, M. McDonald$^{3}$,
    John~T.~Whelan$^{4}$,
    Gustavo Bruzual$^{5}$ \\
  $^{1}$ Max-Planck-Institut f\"ur Gravitationsphysik
  (Albert-Einstein-Institut), D-30167 Hannover, Germany \\
  $^{2}$ Rochester Institute of Technology, 54 Lomb Memorial Drive
  Rochester, NY, USA 14623 \\
  $^{3}$ MIT Kavli Institute for Astrophysics and Space Research\\
  $^{4}$ School of Mathematical Sciences and Center for Computational
  Relativity and Gravitation,
  Rochester Institute of Technology, Rochester, NY 14623, USA  \\
  $^{5}$ Instituto de Radioastronom\'ia y
  Astrof\'isica, UNAM, Campus Morelia, C.P. 58089, Morelia, M\'exico\\
}

\date{Received/Accepted}

\maketitle

\begin{abstract}
  Star formation in galaxies at the center of cooling-flow galaxy
  clusters is an important phenomenon in the context of formation and
  evolution of massive galaxies in the Universe. Yet, star formation
  rates~(SFRs) in such systems continue to be elusive. We use our
  Bayesian-motivated spectral energy distribution~(SED)-fitting code,
  {\bc}, to estimate the plausible SFR values in the brightest cluster
  galaxy of a massive, X-ray luminous galaxy cluster, Phoenix.
  Previous studies of Phoenix have resulted in the highest measurement
  of SFR for any galaxy, with the estimates reaching up to
  $1000~\mpy$. However, a very small number of models have been
  considered in those studies. {\bc} allows us to probe a large
  parameter space. We consider two models for star formation history,
  instantaneous bursts and continuous star formation, a wide range of
  ages for the old and the young stellar population, along with other
  discrete parameters, such as the initial mass function,
  metallicities, internal extinction and extinction law. We find that
  in the absence of any prior except that the maximum cooling rate 
  $< 3000~\mpy$, the SFR lies in the range~($2230-2890$)$~\mpy$.  If we
  impose an observational prior on the internal extinction,
  E(B-V)$ \le 0.6$, the best-fit SFR lies in~($454-494$)~$\mpy$, and we
  consider this as the most probable range of SFR values for
  Phoenix. The SFR dependance on the extinction is a reflection of the
  standard age-extinction degeneracy, which can be overcome by using a
  prior on one of the two
  quantities in question.\\


\end{abstract}

\section{Introduction}
\label{intro}

The last decade has witnessed tremendous progress in the field of
cooling flows in galaxy clusters, both observationally and
theoretically. Observationally, there seems to be a sharp threshold in
cooling time \citep[e.g][]{PaperIII} or equivalently central entropy
\citep[e.g.][]{Cavagnolo2009} below which there is clear evidence of
cooling (1) in the form of cold gas, such as $\ha$ filaments
\citep{Heckman1989,Crawford1999,Conselice2001,McDonald2012a}, CO
emission \citep{Edge2001,Edge2002,Salome2008}, FIR emission
\citep{Edge2010a,Edge2010b,Mittal2011a,Mittal2012,Rawle2012}, (2) in
the form of presence of an AGN \citep[e.g.][]{Mittal2009} and (3) in
the form of star formation
\citep[e.g.][]{Hicks2005,ODea2008,McDonald2011b,Mittal2015,Tremblay2015}. These
observations can now be understood in terms of a particular quantity
first introduced by \cite{Sharma2012}. Their numerical simulations
show that formation of multi-phase filaments depend upon the ratio,
$r=t_{\st {TI}}/t_{\st{ff}}$, of the growth time of the thermal
instability to the free-fall time. When this ratio falls below a
certain threshold, the local thermal instabilities cause the
temperature and density fluctuations to become non-linear and the gas
quickly cools. The cold gas decouples from the hot intracluster
medium~(ICM) and produces spatially-extended ${\ha}$ filaments. These
cold clumps can either rain down on the supermassive black hole or
contribute to star formation \citep[e.g.][]{Gaspari2012,Li2015}. It is
this cold accretion that triggers AGN activity that heats up the
cluster atmospheres. The AGN heating eventually drives this ratio to
go above 10, which halts any further precipitation and subsequently
the AGN activity. This marks the beginning of another cycle fo ICM
cooling, followed by star formation and AGN-heating.

Despite the aforementioned observational and theoretical progress,
star formation rates~(SFRs) in the brightest cluster galaxy of such
cool-core systems are still very poorly constrained with an
unacceptably wide discrepancy between different estimates, and are an
impediment preventing development of a coherent narrative. The chief
limitation of the existing methods of determining star formation rates
(based on the measurement of emission lines, such as $\ha$, or
FUV$-$NUV colour or infrared luminosity) is that there exist {\it
  degeneracies} in that different combinations of the model parameters
can result in the same observed spectral energy
distribution~(SED). This is the underlying reason for the widely
discrepant estimates of star formation rates existing in the
literature. As an example, \cite{ODea2010} used FUV emission to
estimate an SFR of $12~\mpy$ in the BCG of A~1835 on the one extreme
(the study does not mention any errorbars) and \cite{Hicks2005} used
UV-excess to estimate an SFR of $(226\pm9)~\mpy$ on the other. Such a
dispersion in results is in most part due to the different assumptions
made about the star formation history, particularly the age and mass
of the young stellar population.

A Bayesian approach is optimal for such a situation since, given a
prior probability of the parameters (uniform in the simplest case), it
considers all possible combinations of parameters and provides
statistically-derived probability distributions for each parameter.
In \cite{Mittal2015}, we studied the star formation histories of the
brightest cluster galaxies in 10 cool-core galaxy clusters (hereon we
refer to a brightest cluster galaxy in a cool-core galaxy cluster as
``cool-core BCGs''). We used a Bayesian-motivated SED-fitting model,
{\bc}, wherein we let the physical parameters vary over realistic
ranges, and implement marginalization technique to get posterior
probability distributions for the model parameters and quantities
derived from them (such as SFRs). As shown by
\cite{Conroy2010,Pforr2012,Walcher2011}, integrated light from
galaxies at points of the SED sampled well enough can be used to
constrain the basic parameters, provided that marginalization
techniques are used to incorporate the uncertainties in the model.

Constraining the star formation history is imperative for cooling-flow
clusters to be better able to link the cooling of the ICM, star
formation and active galactic nuclei~(AGN)-regulated feedback. In
\cite{Mittal2015}, we find that 9 out of 10 BCGs have been
experiencing starbursts since 6~Gyr ago. While 4 out of 9 BCGs seem to
require continuous star formation, 5 out of 9 seem to require periodic
star formation on intervals ranging from 20~Myr to 200~Myr. This time
scale is similar to the cooling-time of the intracluster gas in the
very central ($<5~$kpc) regions of BCGs. These results are just the
tip of the ice-berg within the paradigm, where multiple epochs of
cooling, star formation and AGN heating are expected to occur.

Similarly, detailed information on the star formation history of a
galaxy also allows us to establish whether or not there is a link
between star formation and AGN heating. In \cite{Mittal2015}, while we
find no relation between the BCG radio luminosity and SFRs \citep[in
accordance with the results of ][]{Li2015}, 4 out of 5 BCGs requiring
multiple outbursts have the highest radio luminosities. This agrees
with the theoretical model of \cite{Sharma2012}, where once the gas
starts cooling, some of it is clumped into molecular clouds leading to
star formation and some of it serves as fuel for the AGN.

In this paper, we apply {\bc} to estimate the star formation rate in
the brightest galaxy of the Phoenix cluster of galaxies,
SPT-CLJ2344-4243, at a redshift of $z=0.596$ (we refer to the BCG as
simply ``Phoenix'' hereafter).  The Phoenix galaxy cluster, although a
relatively recent discovery by the South Pole Telescope using the
Sunyaev-Zel’dovich (SZ) effect \citep{Williamson2011, McDonald2012b},
has an acclaimed status owing to its high X-ray luminosity,
L$_{(2 - 10) \st{keV}} = 8.2 \times 10^{45}$ erg~s$^{−1}$, and the
largest predicted cooling-flow rate (the X-ray mass deposition rate)
known-to-date of about $(3300\pm200)~\mpy$ in the inner 100~kpc
\citep{McDonald2013a}. It is further host to a massive reservoir of
molecular gas
\citep[$M_{H_2} \sim 2\times10^{10}~\ms$][]{McDonald2014}, and a dusty
type-II quasar \citep[e.g.][]{Ueda2013}.

The aspect that makes this cluster of further particular interest is
the ongoing star formation rate in Phoenix inferred using different
techniques and based on different assumptions, with estimates ranging
from about 400~$\mpy$ to about 1500~$\mpy$ \citep{McDonald2012b}. The
study by \cite{McDonald2013a} used five HST/WFC3 broad-band filters and
presented a UV-derived revised estimate of $(798\pm42)~\mpy$, assuming
a mean internal extinction of $E(B-V)=0.34$. However, more recent
results of \cite{McDonald2015} from an analysis combining the previous
HST/WFC3 broad-band filters along with HST/COS FUV and Gemini-S GMOS
optical spectroscopy suggests an SFR of about $(300-500)~\mpy$. This
range reflects the uncertainty in internal extinction and the
dependence of the result on the {\it only} two star formation history
scenarios considered that give the lowest least-square-fit residuals
$-$ (1) continuous star formation since 15~Myr ago, (2) an
instantaneous burst that occurred 4.5~Myr ago.

The above summary of the efforts made to estimate the SFR in Phoenix
far under-states the detailed analysis and assumptions made at each
step. Estimating star formation rates in galaxies is a task far from
simple. There exist numerous theoretical and empirical relations that
connect observables to a star formation rate. Most of these relations,
if not all, are based on stellar population synthesis~(SPS), where a
star formation history~(SFH) along with an
initial-mass-function~(IMF), a metallicity, an extinction law and
internal reddening is assumed, reflecting a large parameter space. It
is common to use supplementary diagnostics to narrow down the range of
parameters but such diagnostic tools rely on yet another set of
assumptions. For example, the use of Balmer emission lines, assuming
case-A (assuming optically-thin nebula) or -B (assuming
optically-thick nebula) recombination, is standard to constrain the
internal reddening. In {\bc}, we simply relax the various assumptions
and use the underlying SPS spectra directly to determine what the data
are telling us. Bayesian marginalization circumvents the issue of a
large parameter space by considering as many models as computational
resources allow us in obtaining meaningful ranges of parameter values.

Despite the large uncertainty in the current estimates of the SFR, it
is clear that the BCG of Phoenix harbours a massive starburst
($> 100~\mpy$). There are very few cooling-flow clusters with such
high SFRs. In general, the SFRs in known cool-core clusters are lower
than the classical cooling-flow rates by factors of 5 to 100
\citep[e.g.][]{Mittal2015,ODea2008}. However, it may very well be that
this observation is due to the lack of a consideration of models that
lead to higher SFRs, and the likelihood of such
models. \cite{McDonald2015} postulate that the vast supply of cold gas
as seen in Phoenix is a feature of a classical cooling-flow model. The
AGN in Phoenix is unique in that it is indicative of both a strong
radiative (quasar) mode and a strong mechanical (radio) mode
activity. A large star formation rate then implies that the
AGN-feedback has not yet coupled with the cooling of the intracluster
medium. In order to be able to make any robust interpretations, it is
crucial that we first make a determination of a {\it reliable} range
of SFRs considering all possible values of the model parameters, so
that we may better address the order of magnitude discrepancy in mass
deposition rates in cooling-flow models, and the exact processes
regulating star formation in some of the most massive galaxies in the
Universe.

We assume throughout this paper the $\Lambda$CDM concordance Universe,
with $H_0 = 71~h_{71}$~km~s$^{-1}$~Mpc$^{-1}$, $\Omega_{\st m} = 0.27$
and $\Omega_{\Lambda} = 0.73$ \citep{Larson2011,Jarosik2011}. The
Galactic extinction towards the line-of-sight of Phoenix is
$E(B-V)=0.016$, which is small as compared to the internal extinction
(both previously known and that derived in this paper), and hence we
ignore it. Moreover, since the extinction is additive in nature, the
true internal extinction can be derived by subtracting the Galactic
extinction from the total extinction.

\section{Data Acquisition and Analysis}
\label{data}

The data used to create the spectral energy distribution for Phoenix
shown in Figure~\ref{best-fit} are described below.

\subsection{HST WFC3-UVIS: Broad-Band Optical Photometry}

We include in this paper optical broad-band imaging from the
\emph{Hubble Space Telescope} Wide Field Camera 3 (HST
WFC3-UVIS). Data in five bands (F225W, F336W, F475W, F625W, F814W)
were acquired using Director's Discretionary Time (PID 13102, PI:
McDonald) in the 2012. These data were reduced using the standard
STScI pipeline, with cosmic rays removed in each individual frame
using the LA Cosmic software \citep{vanDokkum2001}. A more detailed
description of these data are presented in \cite{McDonald2013a}.

For each filter, we extract the total flux within an aperture roughly
3$^{\prime\prime}$ (20 kpc) in radius, which is large enough to
enclose the majority of the flux from all bands \cite[see Figure~3
from ][]{McDonald2013a}.

\subsection{Gemini GMOS: Optical Spectroscopy}
Optical spectroscopy used in this paper were obtained using the GMOS-S
IFU on Gemini South. These data span (5360--9600)~\AA\ in the observed
frame, corresponding to (3356--6011)~\AA\ in the rest frame. The full
field of view for these data is
9$^{\prime\prime}\times5^{\prime\prime}$. For a full description of
the data acquisition and analysis, the reader is directed to
\cite{McDonald2014}. The optical spectrum of the central galaxy was
extracted in the same aperture as described above for the broad-band
data.

\subsection{HST COS: Ultraviolet Spectroscopy}
Far-UV spectroscopy for the central galaxy in the Phoenix cluster was
obtained using the Cosmic Origins Spectrograph on the \emph{Hubble
  Space Telescope} (HST-COS; ID 13456, PI: McDonald). The data used in
this paper consist of two aperture spectra, each having a
2.5$^{\prime\prime}$ diameter. The combined footprint for these
apertures is roughly $4^{\prime\prime}\times2.5^{\prime\prime}$. The
relative positioning of the two pointings yields roughly 80\%--100\%
throughput over the full region for which there is bright UV emission
(i.e., the galaxy center). The full details of these observations and
their reduction are presented in \cite{McDonald2015}.

We matched the aperture of the three instruments by extracting
aperture spectra from the GMOS and HST broad-band filters and matching
the COS throughput as a function of radius within that aperture. While
best efforts were made to ensure that the three data-sets reflected
the same region in the sky, there could still be some remaining
offsets between the instruments. However, we suspect those offsets to
be within the data uncertainties.

\section{AGN and Dust Contamination}
\label{agndust}

X-ray {\it Chandra} data shows that the central 10~kpc is dominated by
a point-source emission in the (2-8)~keV band, which is very likely an
AGN. Diagnostic line ratios suggest a Seyfert-like component in the
nucleus, corresponding to the AGN, and LINER-like component in the
extended regions, corresponding to the cool-core filaments, likely
excited by young stars and shocks \citep{McDonald2012b}. The AGN is
also observed to be radio loud (${\nu}L_{\nu} = 10^{42}$~erg~s$^{-1}$)
\citep{Mauch2003}, albeit with radio luminosity far lower than the
cooling luminosity. The central source, based on the high far-infrared
luminosity, together with the hard X-ray and radio luminosity of the
AGN, is considered to be dusty, highly-obscured. \cite{McDonald2013a}
argue based on further HST UV data, along with the lack of UV emission
along the minor axis of the central galaxy, that the AGN contribution
to the UV luminosity is small ($< 10\%$). In the analysis to follow,
hence, we work under the assumption that the AGN contribution to the
spectral energy distribution of the galaxy may be ignored. The reader
is referred to \cite{McDonald2012b,McDonald2013a} for detailed work on
the AGN in Phoenix and its characteristics, providing justification
for this assumption.

Most observations of Phoenix point to a very dusty system with a
possibly high internal reddening to the extent of $E(B-V)$ being close
to $0.6$.  However, the aim of applying {\bc} is to explore the
stellar population parameters. While we may easily incorporate the
plethora of available infrared data, and thereby increase the number
of constraints, experience tells us that fitting accurate dust
parameters (the number of thermal dust components along with their
masses, temperatures and the dust absorption coefficient) is another
task in itself, which not only entails making assumptions on the dust
model parameters, but also about the background stellar radiation. The
latter defeats the purpose of trying to determine the various stellar
populations and the physical parameters thereof. Furthermore, since
our analysis is based on using a minimum number of assumptions, one of
the requirements becomes to avoid fitting any thermal dust
components. Hence, we restrict our analysis to using data up to an
observed-frame wavelength of around $1~\mu$m.

\section{Bayesian Method}
\label{stats}

Here we describe the details of our code, {\bc}. The basic method was
described in detail in \cite{Mittal2015}; we recap the key features
here.  Note that {\bc} is very similar to iSEDfit
\citep{Moustakas2013} but the two works are independent and differ in
their motivation ({\bc} derives its motivation from the existing
dispersion in the SFR estimates, specifically, in cool-core BCGs).
Using the SEDs generated with integrated flux-densities (using any
available data, broad-band and/or spectroscopic), we fit the data with
a model comprising an old stellar population~(OSP) and a young stellar
population~(YSP), each of which has at least two parameters -- the age
and the total mass. In \cite{Mittal2015}, we hypothesized that the
normalization of the SED corresponding to a synthetically generated
stellar population scales linearly with the total mass in the
stars. Assuming that the mass and the age are the only two parameters,
the flux density at any given frequency, $i$, may be written in the
form, $F_i(M, T) = M\times S_i(T)$, where $M$ is the total mass and
$T$ is the age of the population. $S_i(T)$ is the flux density per
unit mass, which depends on the age. Now, we extend this ansatz to
include other model parameters.

We have a series of flux measurements $\Fis$ with
associated weights $\{w_i\}$, where $1/\sqrt{w_i}$ is the $1\sigma$
uncertainty associated with $F_i$.  Given a family of models $H$
parametrized by YSP mass $\mysp$, OSP mass $\mosp$, and some other
(discretely-sampled) parameters $\lamvec$, if the errors on the $F_i$
are assumed to be independent and Gaussian, Bayes's theorem allows us
to construct a posterior probability distribution
\begin{equation}
  P(\mosp,\mysp,\lamvec|\Fis,H)
  \propto
  P(\mosp,\mysp,\lamvec|H)
  \,\exp\left(-\frac{\chi^2(\mosp,\mysp,\lamvec)}{2}\right) 
\end{equation}
\begin{equation}
   \text{where} \quad \chi^2(\mosp,\mysp,\lamvec)
    = \sum_i w_i
    [F_i-\mosp S^{(\st{o})}_{i}(\lamvec)-\mysp S^{(\st{y})}_{i}(\lamvec)]^2 \, ,
    \label{chisq}
\end{equation}
$S^{(\st{o})}_{i}$ and $S^{(\st{y})}_{i}$ are the flux per unit
mass contributions from the OSP and YSP, respectively. The total
observed flux at a given frequency, $i$, is assumed to be equal to
\begin{eqnarray}
     F_i (\lamvec) &=& F^{(\st{o})}_{i}(\lamvec)+
     F^{(\st{y})}_{i}(\lamvec) =  \mosp S^{(\st{o})}_{i}(\lamvec)+
     \mysp S^{(\st{y})}_{i}(\lamvec) \\\nonumber
                   &=& \mosp S^{(\st{o})}_{i}(\lamvec)+ \mysp \sum_{n=1}^{\nbursts} S^{(\st{y})}_{i,n}(\lamvec)\,
    \label{totalflux}
  \end{eqnarray}
  where $F_i^{(\st{o})}$ and $ F_i^{(\st{y})}$ are the flux
  contributions from the OSP and YSP, respectively, and $\nbursts$ are
  the number of bursts of star forming episodes. $\lamvec$ includes
  the age of the OSP, $\tosp$, and the age of the YSP, $\tysp$. To
  simplify both the approach and the calculations, we assume uniform
  priors, specifically uniform density in $\mysp$ and $\mosp$, with
  the only restriction being $0<\mysp<\mosp$, and that each of the
  discrete values of extinction, metallicity and $\tosp$, and the
  different possibilities for IMF and extinction law, are
  independently equally likely.  Since we consider multiple-starburst
  models with $\nbursts$ bursts evenly spaced in age from $\tysp$ to
  $\nbursts\tysp$, we consider each of the discrete $(\nbursts,\tysp)$
  combinations sampled to have equal prior probability.

From the posterior probability, we can calculate useful probability
distributions for various variables, marginalized over the others, such as
the posterior probability density for the masses
\begin{equation}
  \label{pdfmy}
  P(\mysp|\Fis,H) = \sum_{\lamvec} \int_{\mysp}^{\infty} d\mosp
  P(\mosp,\mysp,\lamvec|\Fis,H)
\end{equation}
\begin{equation}
P(\mosp|\Fis,H) = \sum_{\lamvec} \int_{0}^{\mosp} d\mysp
  P(\mosp,\mysp,\lamvec|\Fis,H)
\end{equation}
or the posterior probability distribution $P(x|\Fis,H)$ for a
discretely-sampled or categorical variable $x$ which is among the
parameters $\lamvec$, whose value at some $x=x_0$ is
\begin{equation}
  \label{px}
  P(x_0|\Fis,H) = \sum_{\lamvec:\ x=x_0}
  \int_{0}^{\infty} d\mosp
  \int_{0}^{\mosp} d\mysp
  P(\mosp,\mysp,\lamvec|\Fis,H)
\end{equation}
By similar means, posterior probability densities can be constructed
for derived quantities such as the mass ratio $\mysp/\mosp$ and star
formation rate\footnote{Since $\mysp$ is the total mass in the YSP,
  the mass in each starburst is $\mysp/\nbursts$.}
$\mysp/(\nbursts\tysp)$.

\section{Stellar Population Synthesis}
\label{SPS}

\subsection{Model Parameters}
\label{modparam}

In order to create synthetic stellar spectra, we used the publicly
available library of evolutionary stellar population synthesis models,
{\sc galaxev}, released by G.~Bruzual and S.~Charlot
\citep{Bruzual2003}. To use the \cite{Bruzual2003} models in the
present version of {\bc}, we modified the tools provided in the {\sc
  galaxev} package to sample the evolving stellar population every
1~Myr. This allows for a finer sampling of the successive star bursts
in the SSP models than if we use the standard time scale. The modified
software is available upon request from Gustavo Bruzual.

The model parameters and the number of types/values they can assume
are (the range and step-size where relevant are given within
square-brackets):
\begin{itemize}
\item Initial Mass Function, IMF: 3 [Chabrier, Salpeter, Kroupa]
\item Extinction laws: 2 [Galactic, extragalactic]
\item Metallicites, $Z$: 3 [0.4 solar, solar and 2.5 solar]
\item Internal Reddening, $E(B-V)$: 15 [0 to 1.4 in steps of 0.1]
\item YSP age, $\tysp$: 6000 [1~Myr to 6~Gyr in steps of 1~Myr]
\item OSP age, $\tosp$: 7 [3~Gyr to 6~Gyr in steps of 0.5~Gyr]
\item Star Formation History, SFH: 2 [continuous star formation,
  instantaneous burst]
\end{itemize}
We terminate both the OSP and YSP age at 6~Gyr because of the maximum
allowed age for Phoenix equal to the age of the Universe at the
formation redshift of the OSP, assumed to be $z_{\st f}=3$, minus its
light-travel time. The detailed description of the model parameters
can be found in \citep{Mittal2015}. The only differences between the
present study and that described in \cite{Mittal2015} are that in the
current study (1) the spacing between bursts is smaller (10~Myr vs
1~Myr) and (2) we consider both continuous star formation (CSF) as
well as instantaneous bursts (or simple stellar populations, SSP)
whereas in \cite{Mittal2015} we considered only the latter. While both
CSF and SSP star formation history models follow
equation~\ref{totalflux}, only the SSP models have $\nbursts$ other
than 1.

Our model allows the possibility of the YSP components to begin as
early as $z\sim 3$ since several studies
\citep[e.g.][]{Eisenhardt2008,Mei2009,Mancone2010,Brodwin2013,Alberts2014}
indicate a model in which the stellar component in galaxies in cluster
centers may have formed as a result of multiple bursts of vigorous
star formation at redshifts as recent as $z\sim1.5$. Furthermore,
there is no theoretical argument why stellar bursts could not have
occurred right after the massive old stellar population was assembled
through mergers. For Phoenix, however, there is no need as such to
impose this requirement since the spectral energy distribution seems
to strongly indicate a star formation history based on a single, rather
young, stellar population. (Section~\ref{results}).

The expected cooling-flow rate, which essentially includes the mass of
the total gas that is available to be churned into stars, and the
shortest time scale over which this can happen (since the gas must
cool down to very low temperatures, $<10~$K to form stars), sets the
upper limit to the star formation rate. The expected cooling-flow
value on the scale of the galaxy~(20~kpc) is about (2000 to
3000)~$\mpy$, and we use this as a prior for our simulations.

\subsection{Data Uncertainties}
\label{dataeb}

As mentioned in \ref{data}, we use both imaging and spectral data for
the SED. An important consideration to bear in mind are the weights
assigned to the data from each category. The overall errorbars (random
as well as absolute) associated with both the imaging and spectral
data are typically $(10-20)\%$. However, the imaging data at
$2371~\AA$, $3353~\AA$ and $4770~\AA$ are sparsely populated in
wavelength. The spectral resolution of HST/COS FUV data is about
$1~\AA/$pix and that of Gemini-S GMOS is about $4~\AA/$pix. In
comparison, the root-mean-square bandwidths of the HST/WFC3 filters
F225W, F336W and F475W are $177~\AA$, $158AA$ and $421~\AA$,
respectively. So that the spectral data (densely populated) and these
three broad-band datapoints are considered at the same footing during
the fitting, we increased the weights associated with the latter by a
factor of about 50, corresponding to a decrease in their errorbars by
a factor of about 7 (since $w_i = 1/\sigma^2$), i.e. reducing the
uncertainty of the (2000-5000)$~\AA$ broad-band data from 15\% to
2.5\%. However, the final results do not seem to be sensitive to the
errorbars associated with the (2000-5000)$~\AA$ broad-band data in
that the most probably values of the model parameters do not change
significantly as the errorbars are reduced from 15\% to 2.5\%.

\subsection{Results}
\label{results}


\begin{figure*}
  \begin{minipage}{0.33\textwidth}
    \centering
    \includegraphics[width=\textwidth]{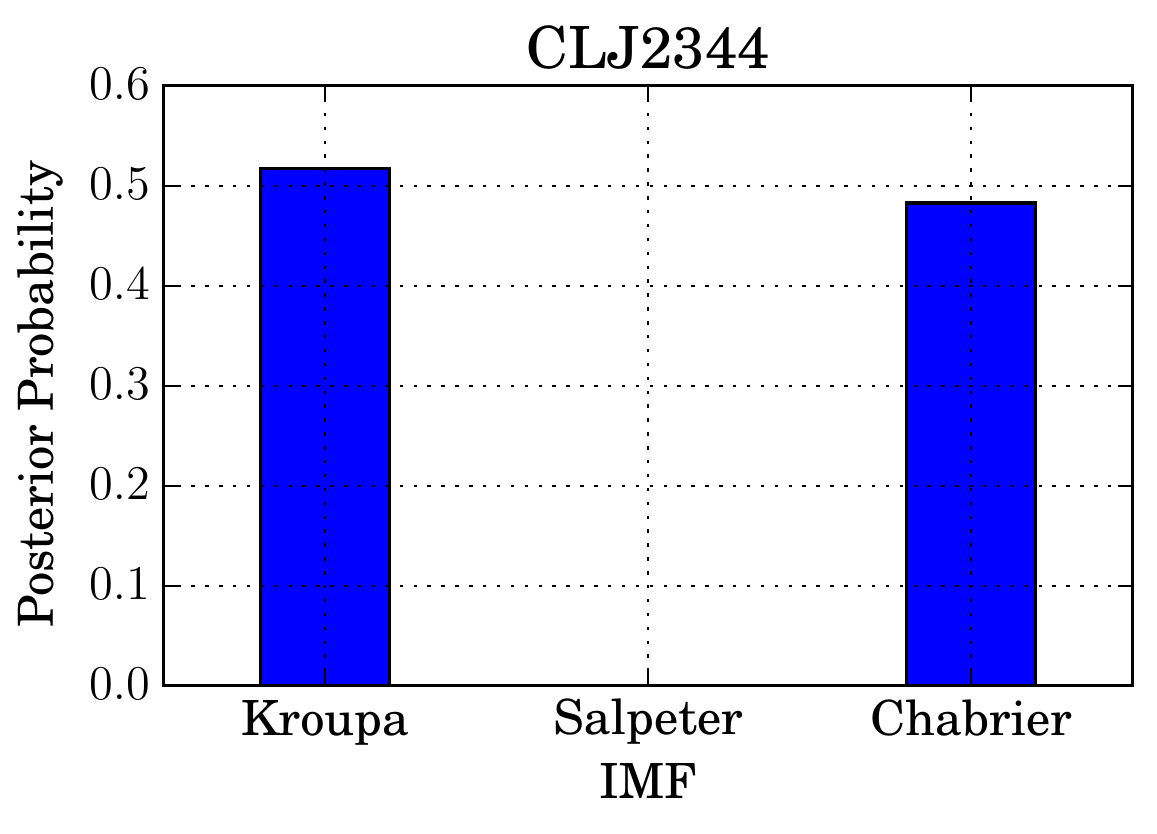}
  \end{minipage}%
  \begin{minipage}{0.33\textwidth}
    \centering
    \includegraphics[width=\textwidth]{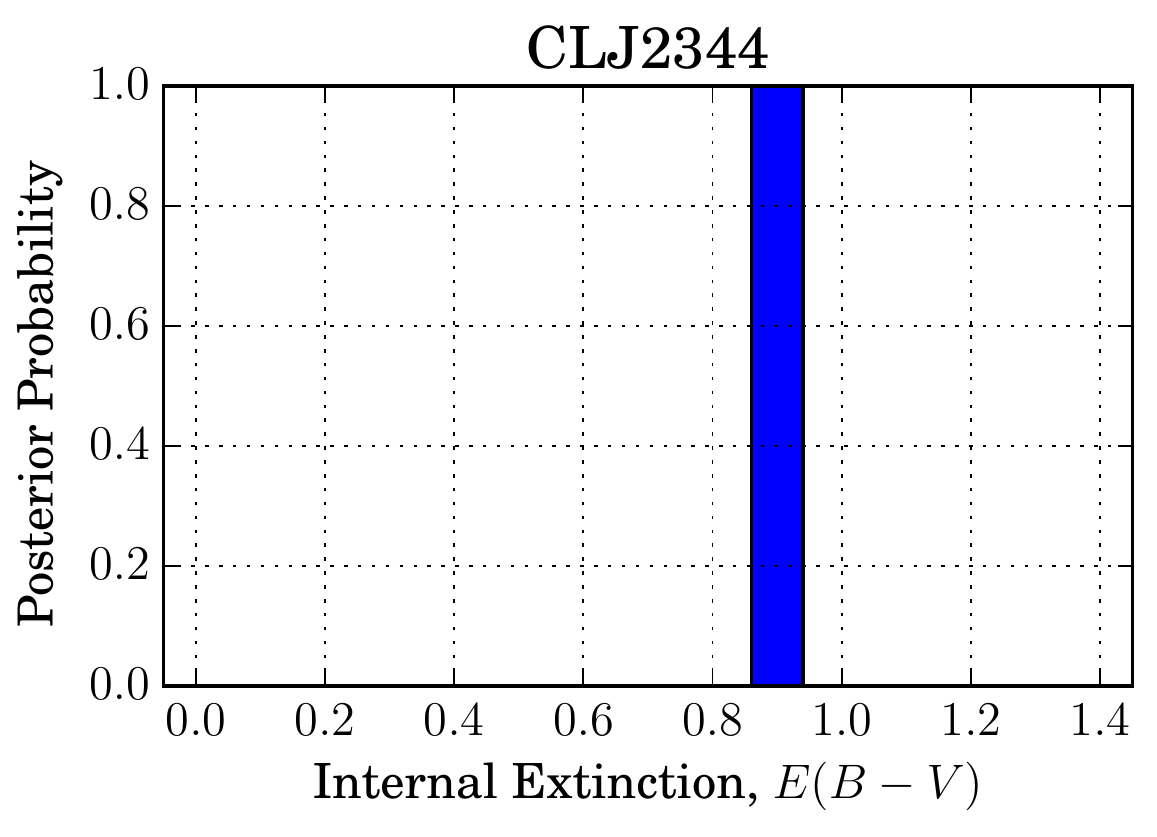}
  \end{minipage}%
  \begin{minipage}{0.33\textwidth}
    \centering
    \includegraphics[width=\textwidth]{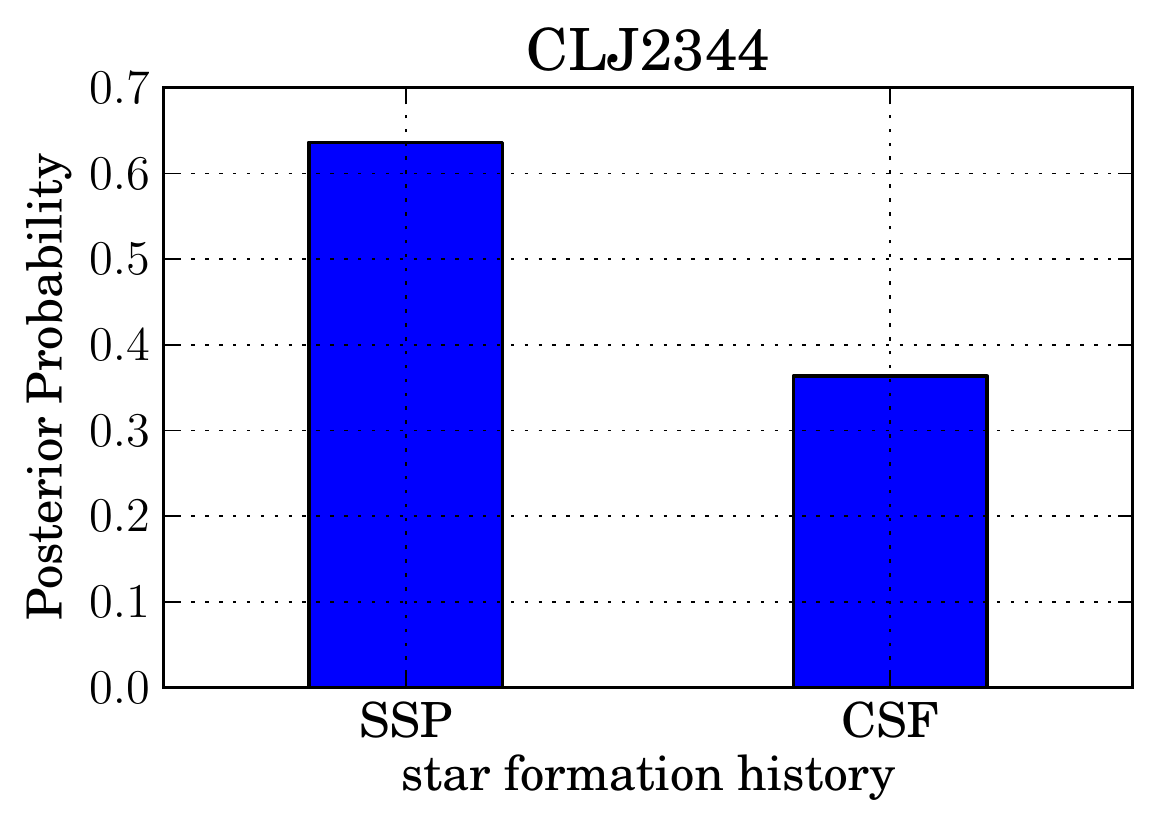}
  \end{minipage}\\
  \begin{minipage}{0.33\textwidth}
    \centering
     \vspace*{-.4cm}
     \hspace*{-0.2cm}
    \includegraphics[width=1.1\textwidth]{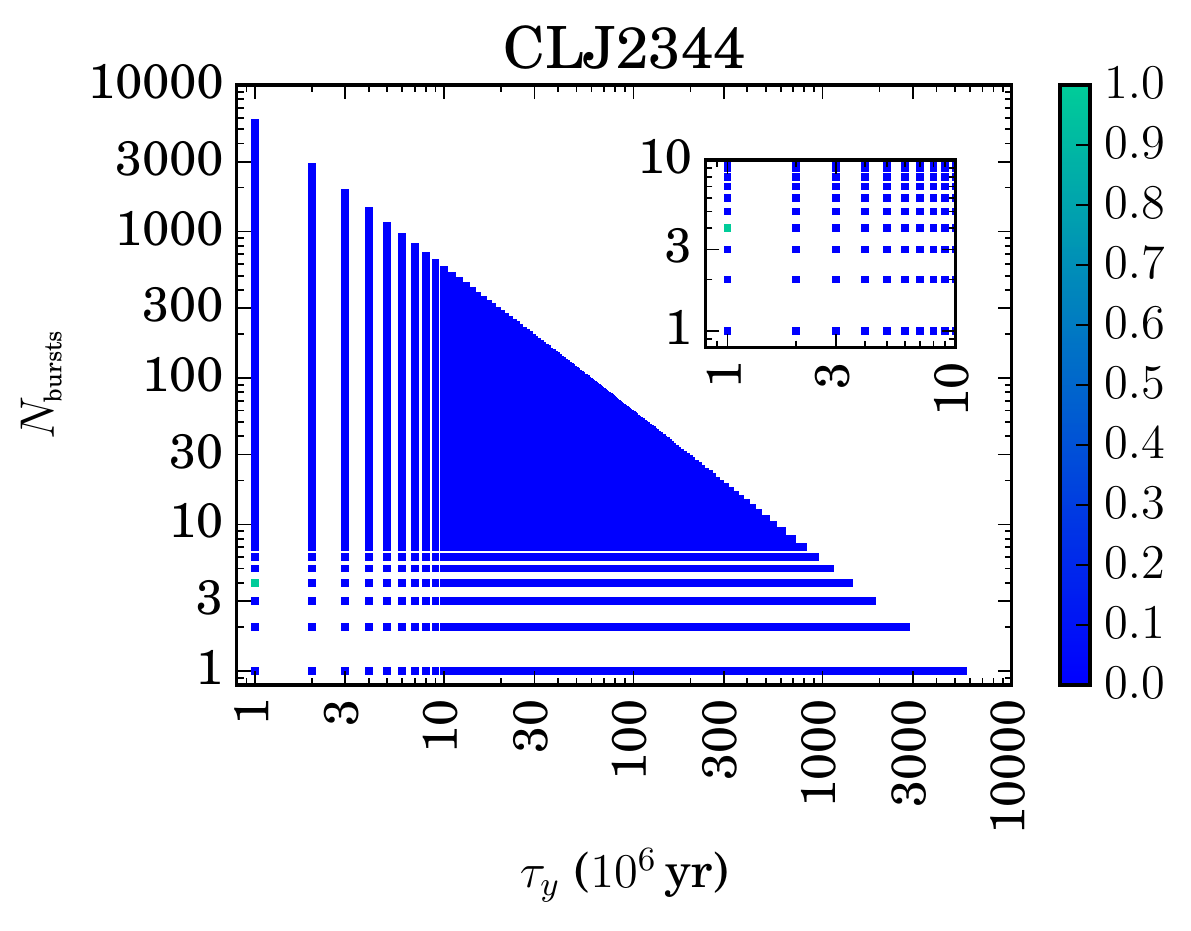}
  \end{minipage}%
  \begin{minipage}{0.33\textwidth}
    \centering
     \vspace*{-1cm}
     \hspace*{-0.2cm}
    \includegraphics[width=1.05\textwidth]{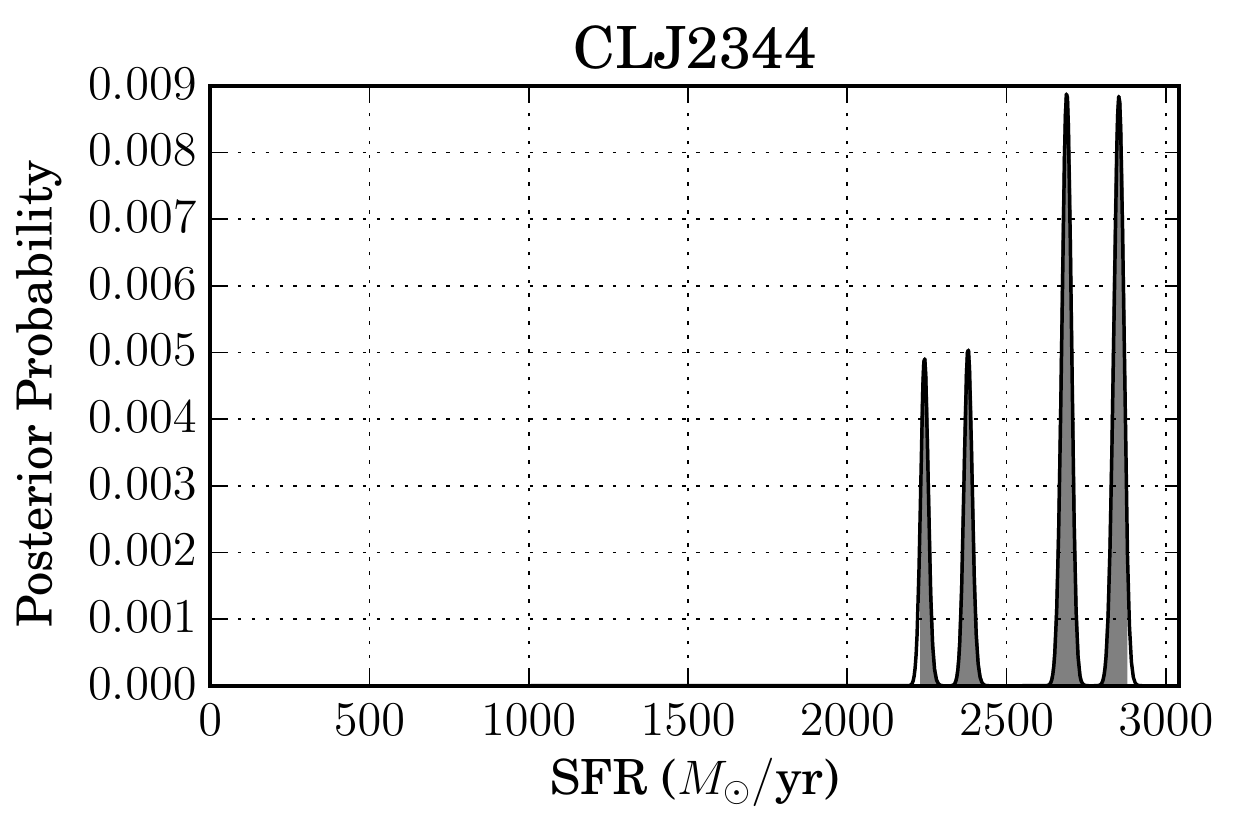}
  \end{minipage}%
  \begin{minipage}{0.33\textwidth}
    \centering
    \includegraphics[width=\textwidth]{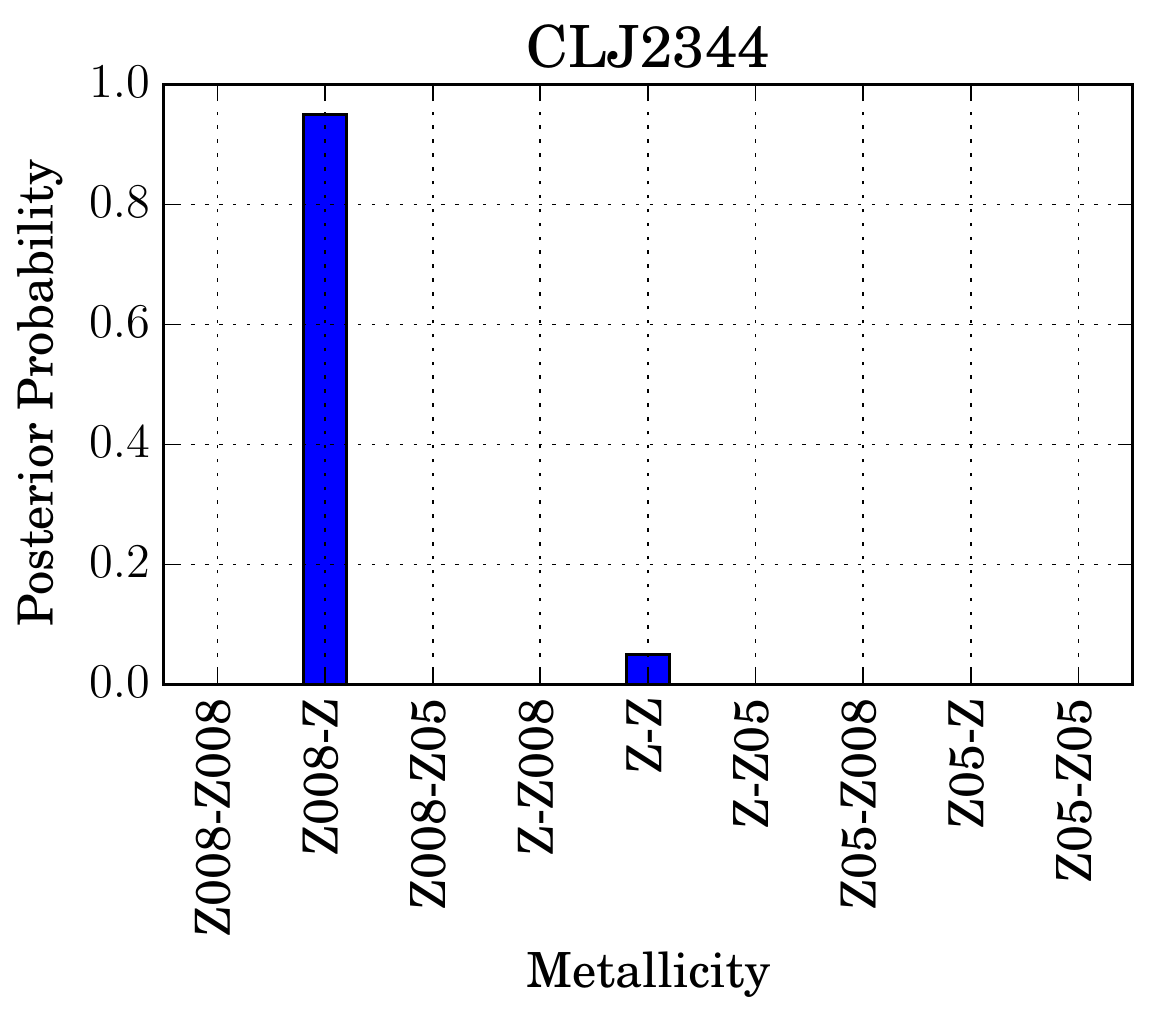}
  \end{minipage}\\
   \begin{minipage}{0.33\textwidth}
     \centering
     \vspace*{-0.2cm}
     \includegraphics[width=\textwidth]{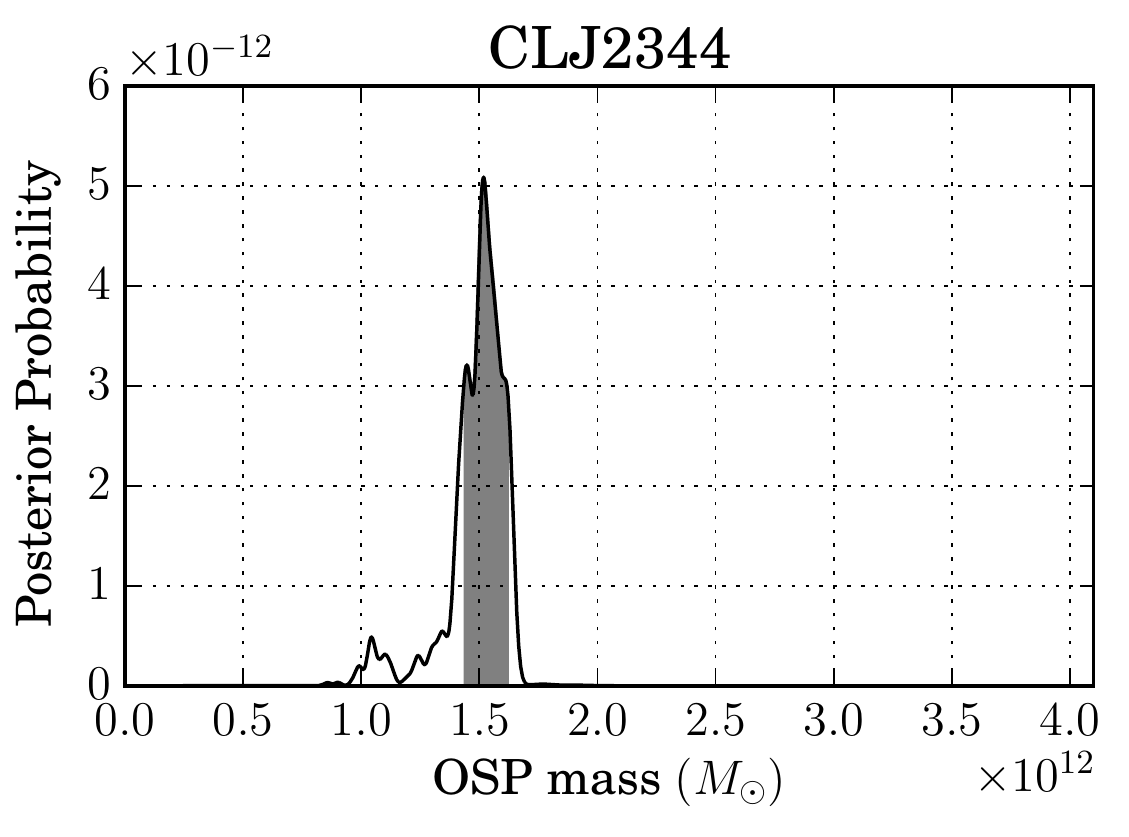}
   \end{minipage}%
   \begin{minipage}{0.33\textwidth}
     \centering
     \vspace*{-0.2cm}
     \includegraphics[width=\textwidth]{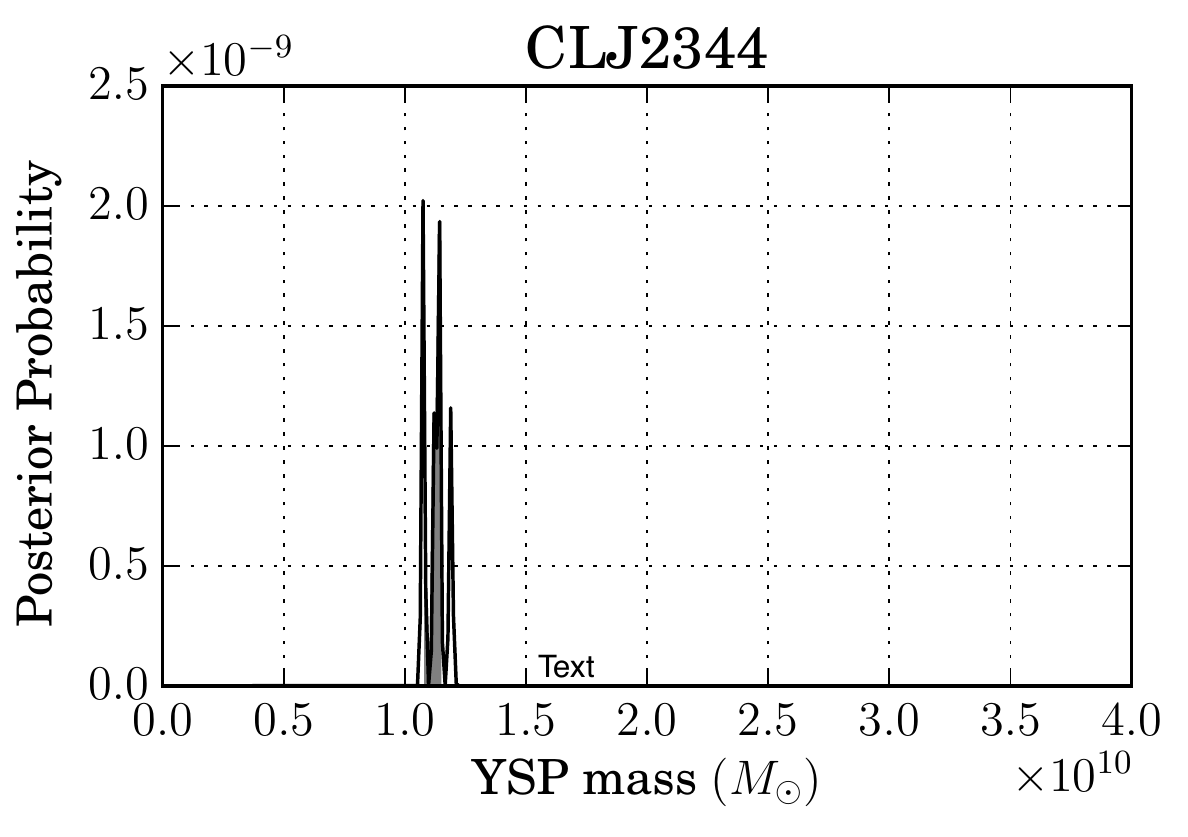}
   \end{minipage}%
   \begin{minipage}{0.33\textwidth}
     \centering
     \vspace*{-0.2cm}
     \includegraphics[width=1.05\textwidth]{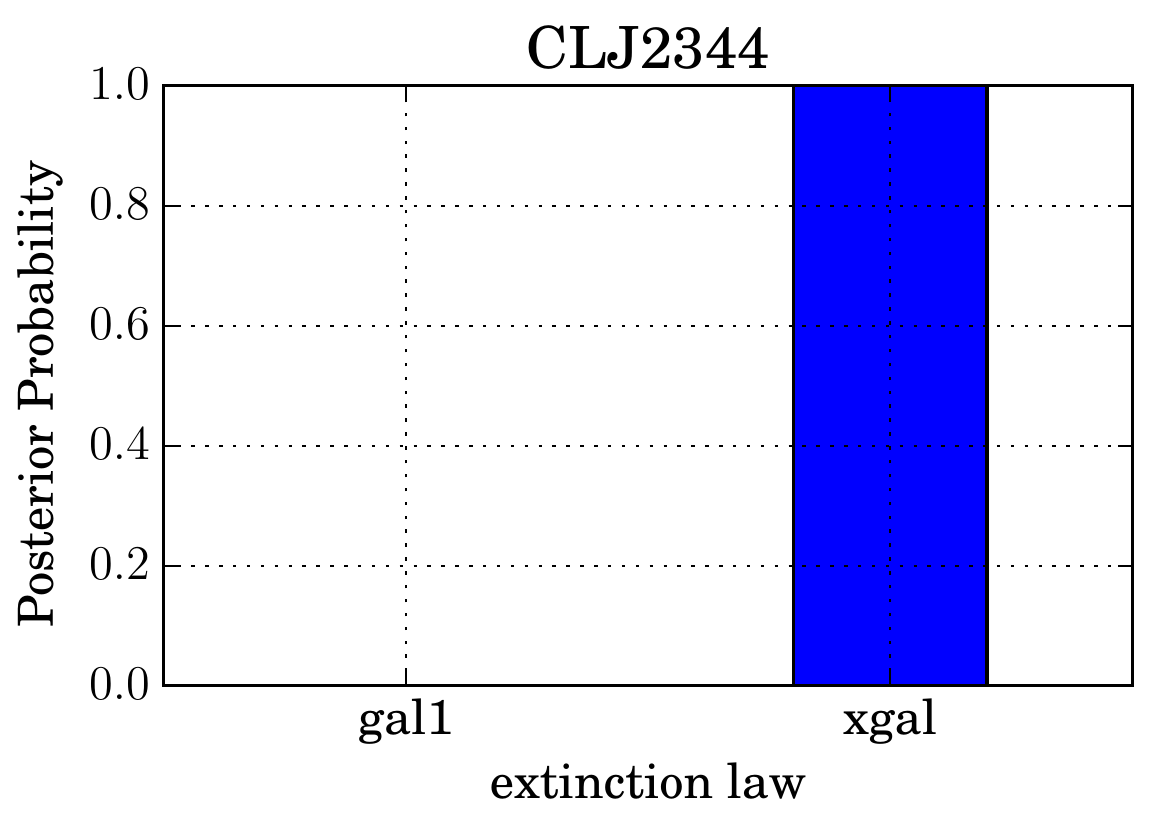}
   \end{minipage}\\
   \begin{minipage}{0.33\textwidth}
     \centering
     \includegraphics[width=1.05\textwidth]{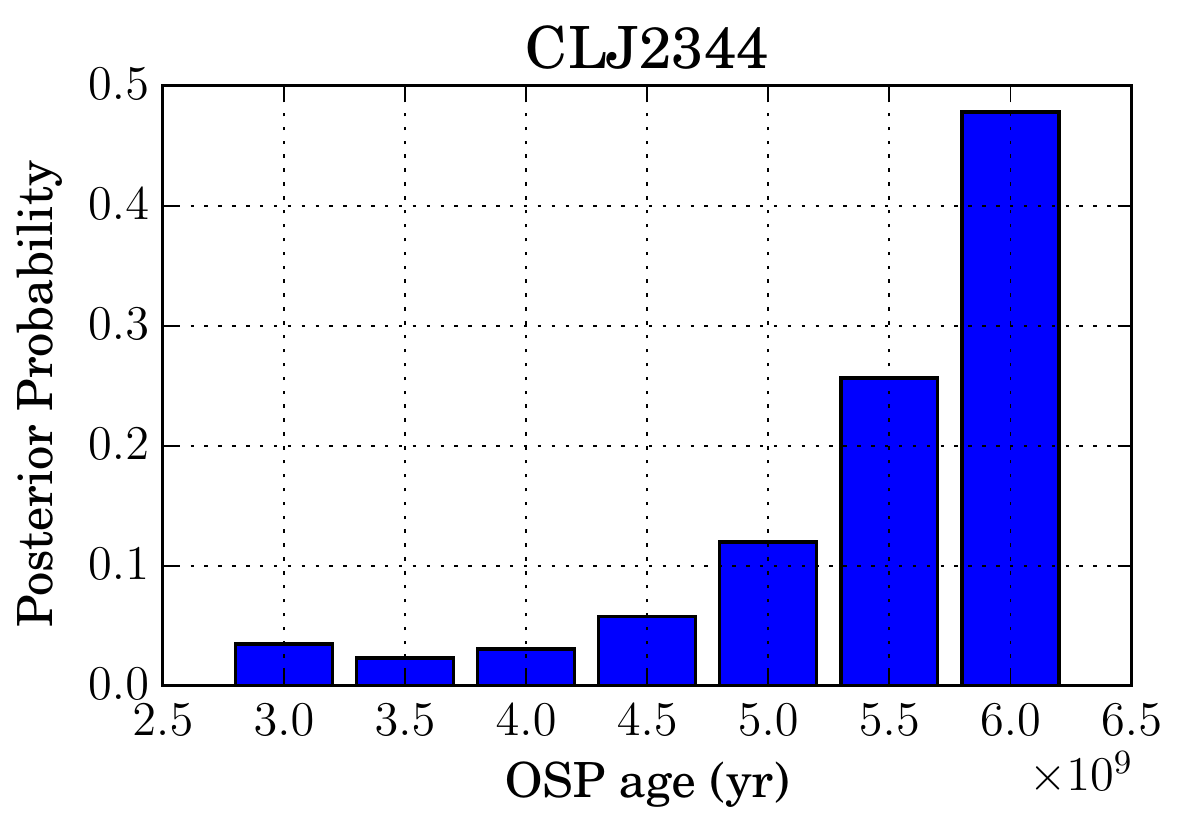}\\
   \end{minipage}%
   \hspace*{0.2cm}
   \begin{minipage}{0.33\textwidth}
     \centering
     \includegraphics[width=1.15\textwidth]{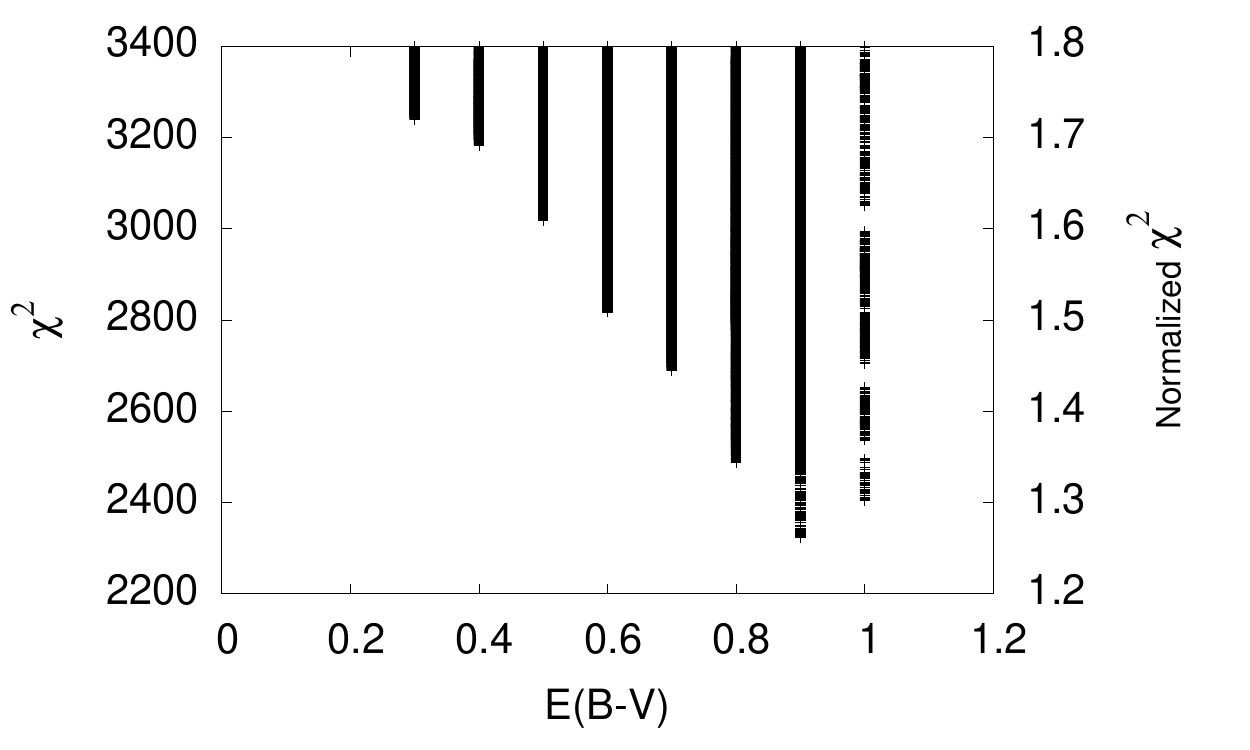}
   \end{minipage}%
   \caption{Posterior probability distributions for the various model
     parameters using the prior $SFR<3000~\mpy$ -- {\it first row
       (left):} initial mass function, {\it first row (middle):}
     internal extinction, {\it first row (right):} star formation
     history, {\it second row (left):} $\nbursts$ vs spacing between
     the bursts (note that in this case significant posterior support
     exists only for the model with $\nbursts=4$ and $\tysp=1~$Myr),
     {\it second row (middle):} star formation rate, {\it second row
       (right):} metallicity, {\it third row (left):} OSP mass, {\it
       third row (middle):} YSP mass, {\it third row (right):}
     extinction law, {\it bottom row (left):} OSP age and {\it bottom
       row (right):} $\chi^2$ as a function of internal extinction.}
   \label{posterior}
 \end{figure*}

We show in Figure~\ref{posterior} the posterior probability
distributions for all the model parameters, where we allowed them to
vary over the respective ranges defined in Section~\ref{modparam},
with the only restriction being that the posterior probability
distribution for the SFR does not exceed 3000~$\mpy$. 

While the data clearly rule out the Salpeter-type IMF, they are not
able to distinguish between the Kroupa- and Chabrier-type (top left
panel). In fact, both the IMFs have about 50\% likelihood of fitting
the data. This is not surprising given the similarities between the
Chabrier and the Kroupa IMFs at the low mass end. The main difference
between Salpeter-type IMF and the other two is that the former
predicts a higher number of low-mass stars.

\subsubsection{Internal Extinction} The internal extinction (shown in
the top middle panel of Figure~\ref{posterior}), on the other hand,
has a single peak at $E(B-V)=0.9$. This value is larger than the
previous measurements made using the Balmer line ratios implying
$E(B-V)$ in the range between 0.3 and 0.4 \citep{McDonald2012b}. In a
later study, \cite{McDonald2013a} published a 2D reddening map, where
the authors deduced reddening values using HST F336W and F475W images,
assuming a flat SED in the absence of reddening
\citep{Kennicutt1998}. Their results based on this assumption show
that $E(B-V)$ ranges from small values ($<0.1$) away from the outer
star-forming regions to 0.6 at the very center. Our results show that
the most plausible value of the average extinction in Phoenix is
higher than any of the previous measurements.

\begin{table*}
  \caption{The most likely physical parameters of the stellar
    populations in the BCG of the Phoenix galaxy cluster, assuming the
    SFR $<3000~\mpy$. ``PI''
    refers to the narrowest 68\% plausible interval for all parameters
    except the star formation rate for which it refers to the
    narrowest 98\% plausible interval. $\tyspold$ refers to the age of
    the oldest YSP. It is relevant only in the case $\nbursts>1$. For
    SFH=SSP only, $\tyspold=\nbursts \times \tysp$.}
    \label{spprop}
  \begin{tabular}{|c | c | c | c | c | c | c |}
    \hline
    Parameter & \multicolumn{2}{c|}{All parameters free} & \multicolumn{2}{c|}{Priors:
                                                           SFH=SSP, IMF=Chab.} & \multicolumn{2}{c}{Priors: $E(B-V) \le 0.6$}\\
    \hline
                     & Mean  &  PI & Mean  &  PI   & Mean & PI  \\
    \hline \hline          
    $\mysp$~($10^{10}~\ms$)   & 1.11 & 1.07 $-$ 1.14  & 1.08 & 1.07 $-$ 1.08 & 2.10 & 1.98 $-$ 2.24\\
    $\tyspold$~(Myr)                 & 4.4   & 3.5 $-$ 4.5 & 4 & $-$ & 44.4 & 41.5$-$46.5 \\
    $\mosp$~($10^{11}~\ms$)   & 15.3 & 14.4 $-$ 16.2 & 15.0 & 14.3 $-$ 15.7 & 12.24 & 11.51 $-$ 12.96 \\
    $\tosp$~(Gyr)                      & 6 & & 6 & & 6 &   \\
    SFR                                      & quadrimodal & 2230 $-$ 2870 & 2690 & 2660 $-$ 2720 & bimodal& 454 $-$ 494\\
    SFH                                      & SSP & $\nbursts=4$, $\tysp=$1~Myr   &   $-$ & $\nbursts=4$, $\tysp=$1~Myr & CSF &  \\
    $E(B-V)$                              & 0.9 &  $-$  & 0.9 & $-$ & 0.6 & $-$  \\
    $\zysp$                              & $\zs$ &  & $\zs$ &  & $2.5\zs$&\\  
    $\zosp$                              & $0.4\zs$  &  & $0.4\zs$ & & $0.4\zs$ & \\  
    IMF                                      & Kroupa/Chab. & & $-$ & & Kroupa/Chab. &   \\
    Extinction Law                     & {\sc xgal}  &  & {\sc xgal} & & {\sc xgal} &\\
    \hline
    \multicolumn{7}{p{15.5cm}}{ The model parameters are the mass of the young and old
    stellar population, $\mysp$ and $\mosp$, the age of the young and
    old stellar population, $\tysp$ and $\tosp$, the star formation
    rate, SFR, the internal reddening, $E(B-V)$, the metallicity of
    the young and old stellar population, $\zysp$ and $\zosp$, the
    initial mass function, IMF, and the extinction law.}
  \end{tabular}
\end{table*}

\subsubsection{Star Formation History} The posterior probability
distribution for the star formation history (top right panel) reveals
that simple stellar populations (a series of instantaneous bursts) fit
the data with a somewhat higher likelihood than continuous star
formation (the latter has about 35\% posterior probability
distribution). The $\nbursts$ versus $\tysp$ plot further points to a
scenario that contains a series of four bursts a Myr apart. Note that
a model comprising four bursts a Myr apart is considered a single
model and not four separate models.

In view of the high extinction value and a very young stellar
population, it is not surprising that the star formation rate lies in
the range (2000 to 3000)~$\mpy$ (center right panel). This is, in
fact, consistent with a star formation rate of $\sim 800~\mpy$ as
estimated by \cite{McDonald2013a} assuming an extinction of 0.6. If
the extinction is considered to be 0.9 instead, the
extinction-corrected flux increases by a factor of three to four at an
assumed observed-frame wavelength of 4000~\AA. This amounts to an
equivalent increase in star formation rate using the
\cite{Kennicutt1998} UV-to-SFR conversion. This back-of-envelope
calculation implies a large range of possible SFRs in the absence of
an accurate knowledge of the internal extinction.

According to the posterior probability distribution of the age of the
old stellar population (the lower most panel of
Figure~\ref{posterior}), the likelihood increases from 3.5~Gyr to
5.5~Gyr and then sharply (with $\sim 50\%$ posterior probability) to
6~Gyr, implying that the oldest generation stars came into existence
at $z\sim3$ (or at an even later redshift, a scenario not included in
our simulations).

\begin{figure}
  \centering
  \includegraphics[width=0.5\textwidth]{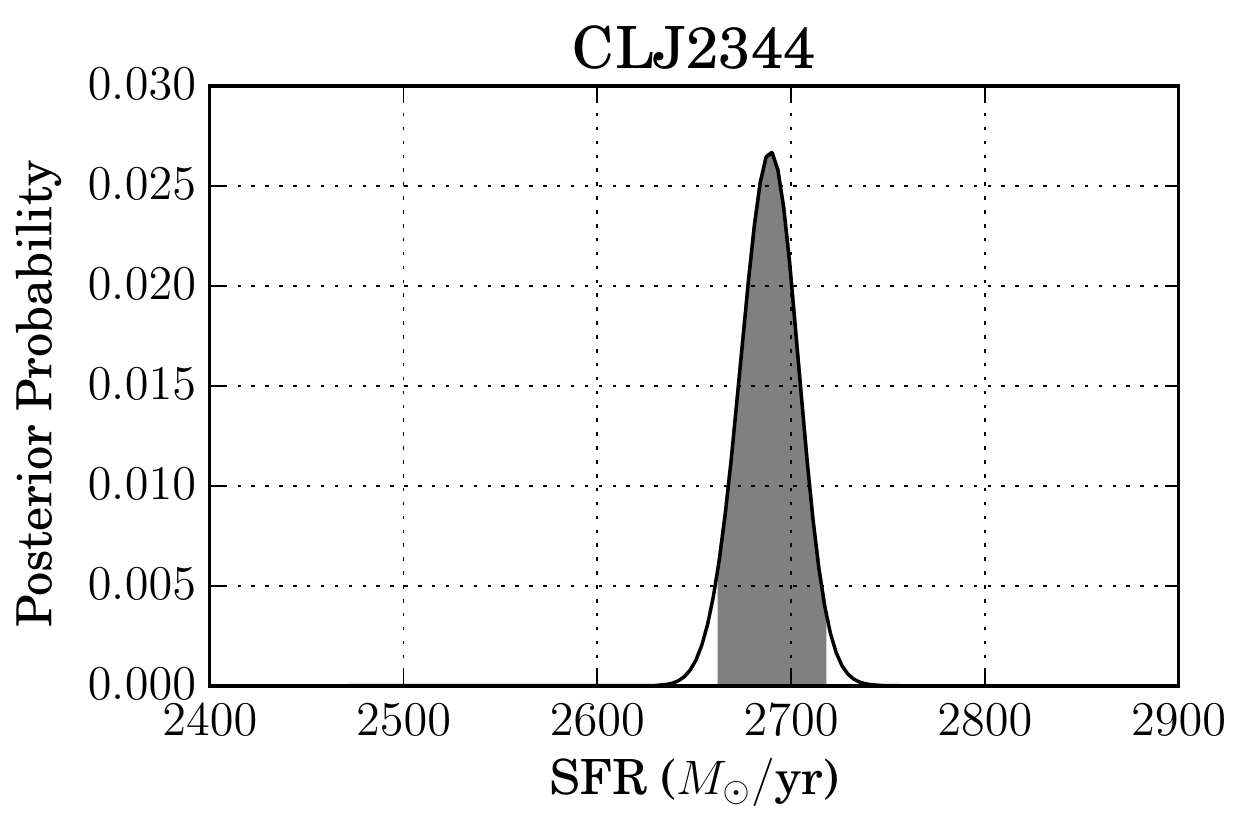}
  \caption{Posterior probability distribution for the star formation
    rate with the initial mass function (IMF) fixed to Chabrier and
    the star formation rate fixed to simple stellar population
    (SSP). There is a single peak at $SFR \sim 2690~\mpy$.}
  \label{chabssp}
\end{figure}

\subsubsection{Star Formation Rate} The posterior probability
distribution of the SFR (center middle panel) is quadrimodal and imply
a star formation rate in the range $2230-2890~\mpy$. The four
different peaks point to the degeneracies in the initial mass function
(Kroupa vs Chabrier) and also the star formation history in
commensurate measure. The two lower peaks correspond to the CSF models
and the two higher peaks correspond to the SSP models. For example, if
the IMF was fixed to Chabrier and the SFH to SSP, the resulting
posterior probability distribution for the SFR will appear as shown in
figure~\ref{chabssp}, with a single clear peak at
$SFR \sim 2690~\mpy$. Similarly, the posterior probability
distribution of the SFR for the case where the extinction, $E(B-V)$,
is limited to 0.6 shows a bimodal distribution due to the degeneracy
in the IMF (the data support only CSF models for the prior,
$E(B-V)<0.6$, hence there are only two peaks).

We tabulate the most likely values of the model parameters in
Table~\ref{spprop} considering three scenarios. The first scenario
corresponds to the best-fit values when all the parameters are allowed
to vary under the restriction that the SFR$< 3000~\mpy$. The second
scenario corresponds to the best-fit values with the added constraints
that the IMF is Chabrier and the SFH is SSP-type. The third scenario
corresponds to the best-fit values when we use the prior,
$E(B-V) < 0.6$, which is the upper limit inferred by
\cite{McDonald2014} using the Balmer emission lines. We quote the mean
and the 68\% plausible interval, the narrowest interval which contains
68\% of the area under the posterior probability density function, for
all parameters except the SFR for which we consider the 95\% plausible
interval.

The star formation rate for the case where the internal extinction is
restricted to be $E(B-V)\le0.6$, as the observations suggest, yields a
star formation rate in the range $(454-494)~\mpy$. We regard the
best-fitting model obtained with $E(B-V)\le 0.6$ as the most reliable
model for Phoenix, and hence the SFRs in the above range as the most
probable values. 

Even though the SFR value of $490~\mpy$ inferred by
\cite{McDonald2015} before making the HST-COS aperture correction lies
in the above range, the results of that work were incorrect by a
factor of $(1+z)$, where $z=0.596$. After correcting the fluxes by
this factor, the star formation rate from the work of
\cite{McDonald2015} is about $780~\mpy$ (assuming other model
parameters do not change). This is higher than what we obtain due to
the fact that while the best-fitting YSP model in \cite{McDonald2015}
corresponds to an instantaneous burst that occurred 4.5~Myr ago, the
best-fitting YSP model from our work (using the prior, $E(B-V)<0.6$)
corresponds to continuous star formation occurring over the past
$\sim 44$~Myr ago. However, \cite{McDonald2015} considered only a few
models ($<100 $) compared to the $\sim 200$ million models considered
in this study. A CSF model with $\tysp=45~$Myr was not tried.

\begin{figure*}
    \centering
    \includegraphics[width=0.9\textwidth]{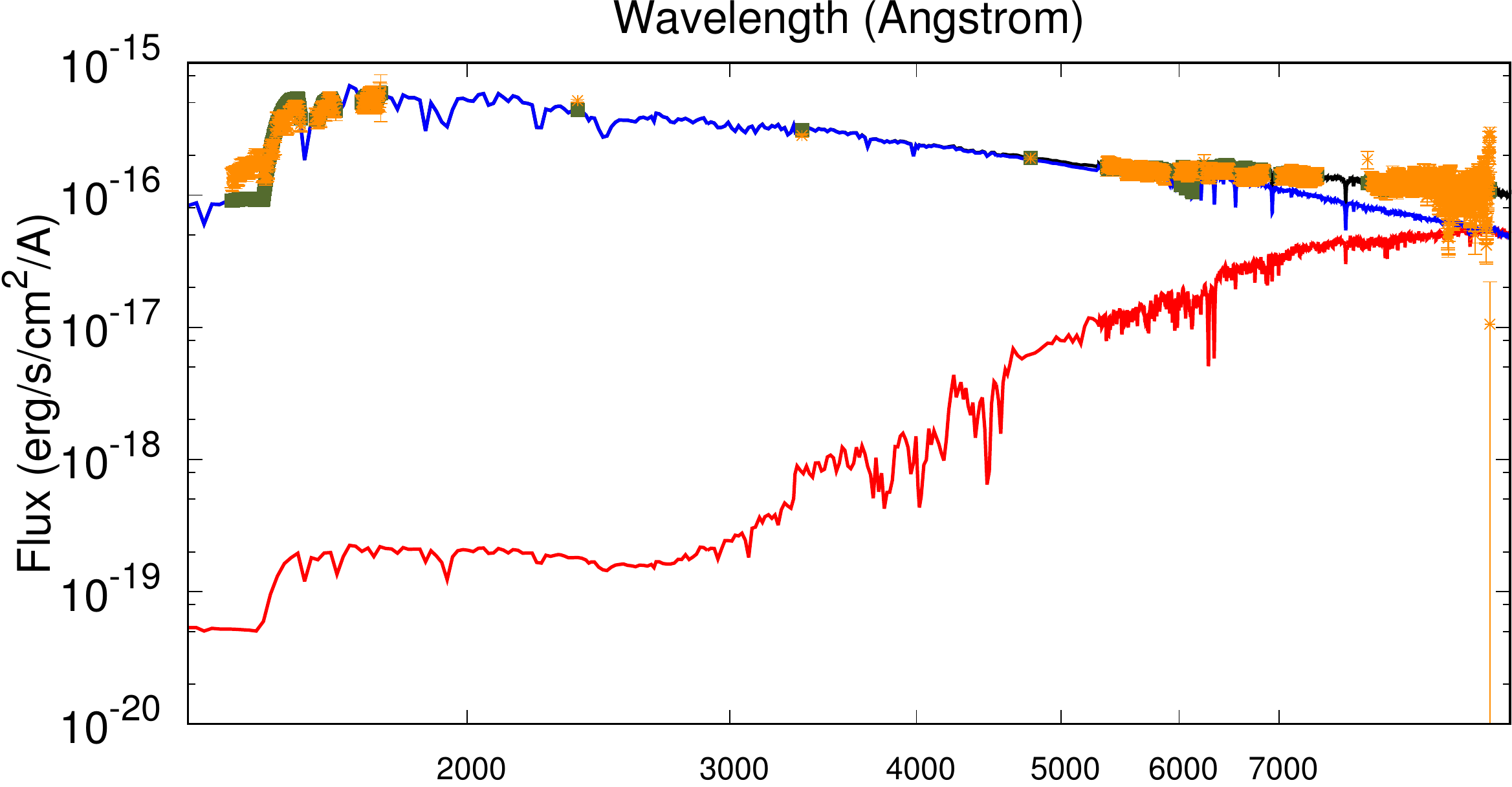}\\
     \includegraphics[width=0.9\textwidth]{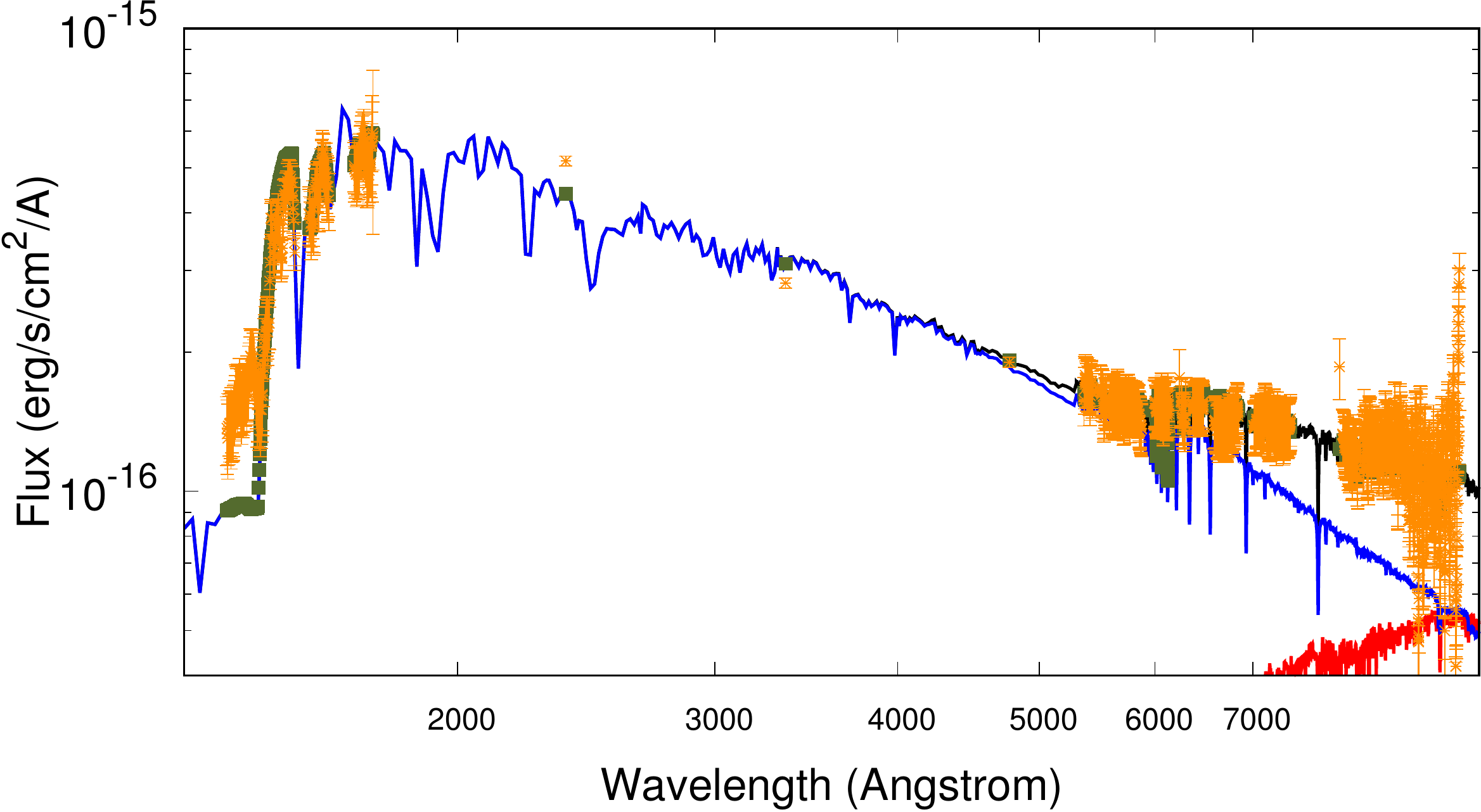}\\
     \caption{An example best-fit plot of the spectral energy
       distribution (observed Galactic-extinction corrected flux vs
       observed-frame wavelength) using the prior $E(B-V)\le0.6$,
       shown by fixing all the discrete parameters and masses to their
       most-likely values, i.e. the modes of their individual marginal
       posteriors. Specifically, the model shown here corresponds to
       SFH=CSF, IMF=Chabrier, $\zysp=2.5\zs$, $\zosp=0.4\zs$,
       extinction law = extragalactic, E(B-V)=0.6 and
       $\tysp=45~$Myr. The red and blue curves correspond to the flux
       contributions from the old and (total) young stellar
       populations, respectively, and the black curve corresponds to
       the total spectrum energy distribution (sum of the old and
       young stellar populations). The green squares correspond to the
       predicted data and the orange crosses correspond to the
       observed data. The bottom panel is a zoom-in of the top panel.}
   \label{best-fit}
 \end{figure*}

 \subsubsection{An example plot} Shown in Figure~\ref{best-fit} in an
 {\it example} best-fit plot created by fixing the model parameters to
 their most likelihood values using the prior $E(B-V)\le 0.6$. Note
 that there are many more models with slightly different combinations
 of model parameters that are capable of reproducing the observed SED
 with the same degree of goodness-of-fit. The errorbars on the most
 likelihood parameters encapsulate all those models and this
 highlights the strength in Bayesian marginalization. The red curve
 depicts the flux contribution by the OSP and the blue curve depicts
 the net-flux contribution by the YSP. The bottom panel of
 Figure~\ref{best-fit} shows the significance of the OSP at
 observed-frame wavelengths beyond 5000~$\AA$ and, therefore, the
 importance of fitting both the old and young stellar component
 simultaneously.

 The best-fit model at the high-frequency end of the HST-COS data
 (1400~$\AA$ to 1500~$\AA$) are not as well explained by the model as
 the rest of the data. This can be due to the fact that our analysis
 includes only two families of models (continuous star formation and a
 superposition of periodically-based bursts). However, we do not
 expect these data to affect significantly our estimates of the
 parameters in the current models since this is a small fraction of
 data points (4\%). In future analyses, we aim to expand our code to
 include multiple stellar populations (non-periodic) and other
 families of SFH, such as exponential-declining star formation,
 stellar bursts of certain length, delayed star formation, all of
 which are available within {\sc galaxev}. Similarly, we aim to work
 with a finer grid of discrete variables (metallicity and also
 evolutionary tracks), which along with a new set of stellar spectra
 designed to model nebular emission in star-forming galaxies,
 particularly over UV and optical range, have recently become
 available within {\sc galaxev}. As an example, a crude exercise shows
 that the break observed around 1450~$\AA$ can be reproduced by adding
 together SEDs corresponding to 6~Myr and 120~Myr stellar populations
 with a metallicity of $Z = 0.017$ and Chabrier IMF.  We also aim to
 modify our code as to allow the parameters of the extinction law vary
 (see Section~\ref{extlaw}). This will affect the highest frequency
 data the most since extinction scales with frequency. Finally, we
 plan to generalize the current model of independent uncertainties
 associated with the data points $\{F_i\}$ with a more general
 framework including correlations among nearby frequencies and/or
 within the spectroscopic sample from each instrument. While we have
 made an attempt to match apertures between different data sources,
 there may still be aperture effects that could lead to slight
 normalization offsets and the above generalization will account for
 this.

\section{Discussion}
\label{discussion}

It is worth considering another scenario where we do not impose the
SFR$<3000~\mpy$ restriction. By considering all models with no
restriction on SFR yields an even higher best-fit extinction,
$E(B-V)=1.1$, and a continuous star formation that commenced 3~Myr ago
with the cumulative mass three times more than that listed in
Table~\ref{spprop}. These properties result in an enormous SFR of
about $(12000-13000)~\mpy$. A recent study conducted by
\cite{Prasad2015} probed the evolution of cool cluster cores in the
presence of bipolar AGN feedback based on hydrodynamic
simulations. Their results show that even though the average SFR is
suppressed by AGN feedback over gigayear timescales, the instantaneous
cold gas mass-inflow rate may be similar or even higher than the
cooling flow value, depending upon the region considered where the gas
condenses profusely (on the scale of a few kpc) and the feedback
efficiency parameter. This is a direct consequence of multi-phase
cooling. Once the multi-phase cooling initiates, the condensation of
gas can proceed at almost the free-fall time. Under these conditions,
the instantaneous SFR timescale may be as short as the dynamical
time. So while a star formation rate in Phoenix much greater than the
cooling-flow rate is physically possible, it is unlikely that we have
caught the system in a state with a likelihood of 5\%-10\%
\cite{Prasad2015}. Moreover, the observations of \cite{McDonald2013a}
do not point to an internal extinction as high as $E(B-V)=1.1$.
Likewise, as discussed in the next Section, SFR estimates from other
independent diagnostics also point to an SFR below $3000~\mpy$. All put
together, there is very little observational evidence for models that
contribute to such a high SFR, and hence, a cut-off of
$SFR <3000~\mpy$ seems justified.

\subsection{Comparison with SFR Estimates from other Diagnostics}
\label{SFRcompare}

Now we compare the SPS-derived SFRs to those derived based on other
diagnostics. A concise summary of all the diagnostics can be found in
\cite{McDonald2012b}. Here we mention the two most important
diagnostics.

First we consider the $\ha$ line emission. Since the $\ha$
observations of \cite{McDonald2012b} are over a smaller aperture than
the $\hb$ observations of \cite{McDonald2014}, we use the $\hb$ line
flux from the latter to calculate the $\ha$ line flux. First we assume
the best-fit value $E(B-V)=0.9$ for the internal extinction and the
extra-Galactic extinction law (with $R_v=4.05$), as preferred by the
data with a 100\% peak marginalized probability. Using the observed
$\hb$ flux of $7.3 \times 10^{-15}$ erg$^{-1}$~s$^{-1}$~cm$^{-2}$, and
assuming the standard case-B (the optical thick limit) recombination
ratio of $\ha/\hb = 2.86$, we derive an extinction-corrected $\ha$
luminosity of $1.63\times 10^{44}~$erg~s$^{-1}$. Using the Kennicutt
$\ha$-to-FIR conversion, which assumes a Salpeter IMF and solar
metallicity \citep{Kennicutt1998}, we derive a $\ha$-based star
formation rate of $1290~\mpy$. In order to convert this value for a
Chabrier or Kroupa IMF, we apply a factor of 1.59 for Chabrier IMF and
1.49 for Kroupa IMF \citep{Madau2014}, and obtain $2048~\mpy$ and
$1925~\mpy$, respectively. Similarly, if $E(B-V)=0.6$ instead, as
favoured by the Balmer line emission ratios, the corresponding SFRs
are $737~\mpy$ for Salpeter, $1170~\mpy$ for Chabrier and $1100~\mpy$
for Kroupa.

Of all the diagnostics, far-infrared~(FIR) emission is usually
considered the most direct and reliable diagnostic for star formation
rate since the FIR photons go through negligible extinction. However,
there are two caveats $-$ (a) FIR emission for Phoenix is not very
easily measurable due to a significant (dominant) contribution due to
the AGN \citep[40\%-86\%][]{McDonald2012b,McDonald2014,Tozzi2015} and
(b) the conversion relations between FIR luminosity and SFR, such as
the Kennicutt FIR-to-SFR relation \citep{Kennicutt1998}, are based on
the assumption that all of the ionizing, UV and optical photons are
absorbed by the dust and reemitted as far-infrared thermal emission
via the photoelectric heating of dust. Under such a scenario, there
should be no detectable Ly-$\alpha$ emission since all Ly$\alpha$
photons emitted by the ionized nebular gas surrounding the hot young
OB stars should be absorbed by the dust \citep{Mas-Hesse1991}. This is
not true for Phoenix, which indeed shows a strong Ly-$\alpha$ line
emission.

Keeping the above caveats in mind, we use the total FIR luminosity
derived by \cite{McDonald2012b},
$L_{\st{ IR}} = 9.5\times10^{12}~\ls$, which resulted from fitting a
single 87~K dust component to {\it Herschel} PACS and SPIRE data.
Assuming a 40\% contribution by the AGN \citep{McDonald2012b} results
in $L_{\st{ IR}} = 3.8\times10^{12}~\ls$ attributed only to stellar
heating and assuming a 86\% contribution by the AGN \citep{Tozzi2015}
results in $L_{\st{ IR}} = 1.3\times10^{12}~\ls$. Now we use the
Kennicutt FIR-to-SFR relation, which assumes a Salpeter-type IMF and
continuous bursts of age (10-100)~Myr, and obtain a star formation
rate of $224~\mpy$ and $654~\mpy$ for 86\% and 40\% AGN contributions,
respectively. Converting these for Chabrier-type and Kroupa-type IMF
implies an SFR of about (355-1040)~$\mpy$ and (335-975)~$\mpy$.

The $\ha$ emission line yields relatively high SFR estimates but these
are subject to internal extinction. The FIR-based SFR estimates, on
the other hand, are much lower and not significantly influenced by
internal extinction, although these should be considered as a lower
limit because of the second caveat mentioned above. Assuming the
FIR-estimates to be closer to the true value implies that the internal
extinction is likely $E(B-V) \le 0.6$, which yields the SFR in the
range (454-494)~$\mpy$. Under this scenario, it may very well be that
a non-negligible fraction of the Balmer line emission has a
non-stellar origin, such as collisional heating and ionization by
secondary electrons within the cold filaments produced by the hot ICM
via reconnection diffusion \citep{Fabian2011}. If true, this will
lower the SFR estimate based on Balmer line emission. The above range
of SFR value also agrees well with the SFR of $(530 \pm 53)~\mpy$ that
\cite{Tozzi2015} obtained from a detailed analysis of the far-infrared
SED.

Lastly, we compare our SFR values to the spectral mass deposition
rates obtained with the {\it XMM-Newton} MOS data
\citep{Tozzi2015}. According to the best-fit value from the single
{\sc mkcflow} model in the (0.3-3.0)~keV temperature range, the mass
deposition rate (MDR) is [620 (-190 + 200)$_{\st {stat}}$ (-50 +
150)$_{\st{ syst}}$]~$\mpy$, once again consistent with the best-fit
range of values for $E(B-V) \le 0.6$.  However, \cite{Tozzi2015} found
the MDRs to be a function of the temperature, with the MDRs ranging
from several 1000~$\mpy$ for (1.8-3.0)~keV data to about 380~$\mpy$
for (0.45-0.9)~keV data. They concluded that the mass deposition rate
found with a single {\sc mkcflow} model (assuming single-temperature
gas) should not be taken as representative of the entire cooling flow,
especially since the gas in Phoenix is observed to cool down to much
lower-than ambient temperatures than other cool-core clusters.

\begin{figure}
  \centering
  \hspace*{-0.85cm}
  \includegraphics[width=0.6\textwidth]{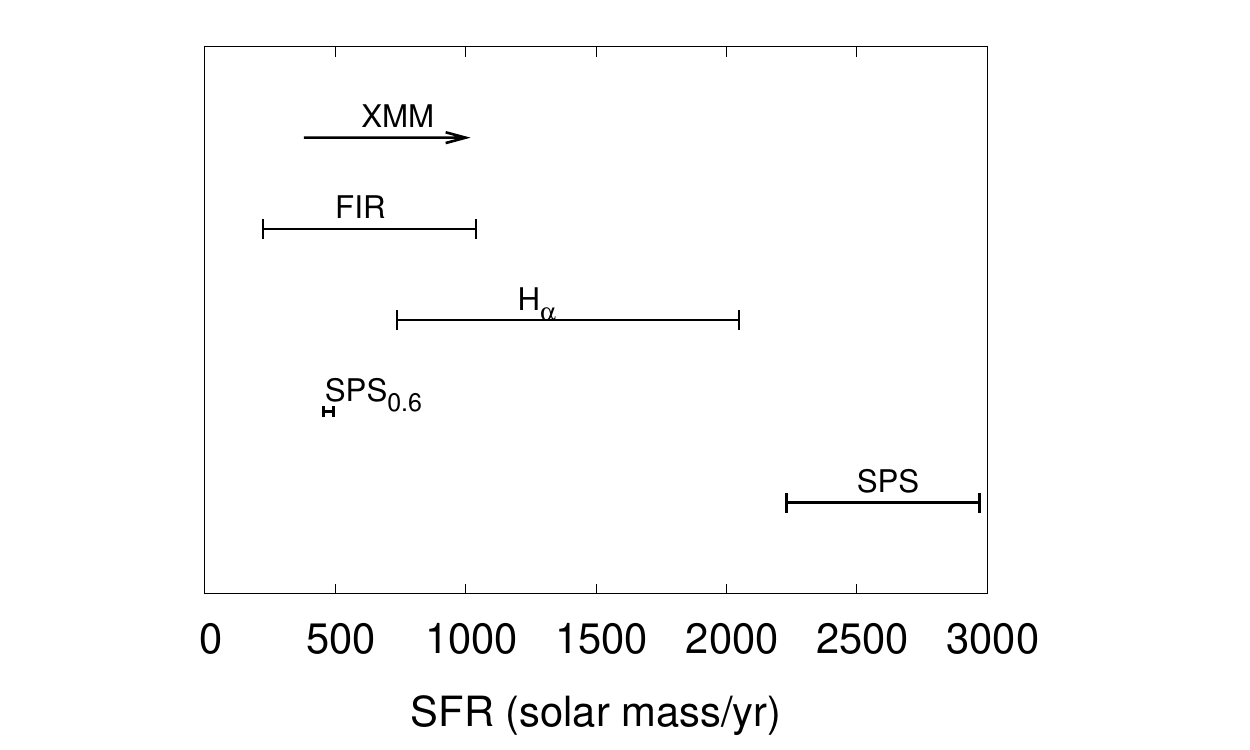}
  \caption{The ranges of SFR values obtained using different
    diagnostics as discussed in Section~\ref{SFRcompare} --
    (a)~stellar population synthesis (SPS) (b)~$\ha$ and
    (c)~far-infrared emission (FIR). We also show the range of
    spectral mass deposition rates as obtained with the {\it
      XMM-Newton} (XMM). We show the 95\% plausible range using SPS
    for both the cases when no prior on extinction is used and when
    the prior E(B-V)$\le 0.6$ (SPS$_{0.6}$) is imposed.}
  \label{sfr-estimates}
\end{figure}

The ranges of star formation rate values as derived from different
diagnostics are summarized in Figure~\ref{sfr-estimates}.

\subsection{Upper Limit to Star Formation}
\label{UL-SFR}

Now we test whether the best-fit young stellar population adheres to
the Eddington limit for star formation \citep[e.g.][]{Scoville2001},
beyond which the stellar radiation pressure on the dust of the stellar
burst will exceed the self-gravity of the molecular cloud, causing the
dust to disperse. The radiation pressure exceeds self-gravity of the
star forming cloud when $(L/M)_{\st {SF}} > 500\ls/\ms$, where the
subscript ``$\st{SF}$'' refers to the star forming cloud.
\cite{McDonald2014} used the ratio of the FIR luminosity,
$L_{\st {IR}}$, to the mass of the molecular hydrogen, $\mh$, to
calculate the luminosity-to-mass ratio of the star forming cloud as
$440\ls/\ms$.

One concern with the results where all the parameters are allowed to
vary, including the extinction, is the high star formation rate of
$2500~\mpy$ that is obtained. Since we have the synthetic spectra of
both the old and the young stellar populations, we can directly
estimate an upper limit to the luminosity-to-mass ratio of the star
forming cloud given the properties of our best-fit YSP. For this, we
calculated the total bolometric luminosity associated with the
cumulative YSP (considering all the four bursts a Myr apart) and
obtained a value of $L=4.1 \times
10^{46}~$erg~s$^{-1}
= 1.08\times10^{13}~\ls$.
Next we considered the mass of the molecular hydrogen, $\mh$, equal to
(2.2$\pm$8.8)$\times 10^{10}~\ms$ as obtained based on the CO(3-2)
observations \citep{McDonald2014}. We can now evaluate a lower limit
and upper limit of luminosity-to-mass ratio of the star forming
site. The lower limit is obtained by considering
$L/[\mysp + (\mh)_{\st {UL}}]$, since the mass of the entire star
forming cloud is expected to contribute to self-gravity.
$(\mh)_{\st {UL}}$ is the upper limit to $\mh$ and $\mysp$ is the mass
of the YSP. The upper limit of $(L/M)_{\st{SF}}$ is obtained by
considering $L/(\mh)_{\st {LL}}$, where we only consider the lower
limit to $\mh$.  Using the above relations, we derive an upper limit
of $(L/M)_{\st{SF}}$ as $490~\ls/\ms$ and a lower limit of
$(L/M)_{\st{SF}}$ as $110~\ls/\ms$, both of which conform to the
Eddington limit of $500~\ls/\ms$. Thus, we do not consider the
properties of the YSP, such as the estimated SFR, even when $E(B-V)$
is allowed to be free, to be a problem within the framework of
self-regulated star formation. In other words, based on the Eddington
limit to star formation, a high value of redenning such as
$E(B-V)=0.9$ that results in a high SFR value can not be ruled out.

\subsection{Self-Regulated Heating and Cooling of the ICM}
\label{UL-SFR}

Next we address how the star formation history fits in together with
the cooling of the intracluster medium and AGN heating. Not
considering any observational priors on the reddening, the SSP models
have a higher posterior probability than the CSF models owing to
slightly lower $\chi^2$ values. The most probable star formation
history model with $E(B-V)=0.9$ points to a series of four bursts
1~Myr apart. A continuous star formation, on the other hand, that has
been onset since 5~Myr ago yields almost similar plausible ranges of
values for the model parameters [e.g. the 98\% plausible interval for
the SFR is $(2220-2270)~\mpy$]. The SEDs for the two cases are
indistinguishable. Therefore we do not make any distinction between
the two scenarios. Our code only proves that, mathematically, a model
with four bursts that are 1~Myr apart fits the data slightly better
than the one with continuous star formation that commenced 5~Myr
ago. The Bayes factor (the ratio of the evidence for the two models,
SSP and CSF) is about two, which implies that the data do not show any
strong preference for one model over the other. On the other hand,
imposing an observational prior $E(B-V)\le0.6$, the SFH seems to
strongly prefer a continuous star formation scenario that has been
going on for the last 45~Myr. The SFH reveals a relatively recent star
formation activity either way.

In \cite{McDonald2015}, the authors claim detection of two pairs of
cavities 20~kpc and 100~kpc apart, where the outer ``ghost-cavities''
have a rather low signal-to-noise, that imply two AGN bursts with a
duty cycle of 100~Myr. Irrespective of the two SFH scenarios (SSP/CSF
over the last 5~Myr or CSF over the last 45~Myr), assuming that a
cooling epoch entails both star formation and blackhole accretion
followed by AGN feedback in the form of outbursts that can be traced
by X-ray cavities, we do not find any evidence of star formation
100~Myr ago. On the other hand, our code {\bc} currently is not
capable of handling complicated star formation histories, such as
those containing multiple young stellar populations with a
non-periodic separation. For example, our code can not model a
composite SED consisting of a 6~Gyr OSP, a 100~Myr YSP and a 5~Myr
YSP. Hence, no conclusions may be drawn to confirm the presence of yet
another pair of outer cavities. As the next step, we will be making
modifications to our code so that such models may be included in the
analysis, data permitting.

Assuming the cool-core in Phoenix formed around the same time as other
cool cores at around a redshift of 1 (or later) \citep{PaperIII}, an
absence of any star formation between 6~Gyr and 45~Myr implies a
heating mechanism that was very effective at regulating the cold gas
in a way that prevented any star formation during that time. Based on
a high X-ray photon luminosity, \cite{McDonald2015} posit that the AGN
in Phoenix is undergoing a transition from ``quasar mode'' to
``radio-mode''. Our results further corroborate this picture with the
quasar-mode as the prime heating mechanism up until recently. As the
next step, it would be interesting to see the results from a
high-resolution radio image of the AGN at the center of Phoenix so as
to investigate the properties of the central AGN, such as its
luminosity and age.

\subsection{Dust Extinction Law}
\label{extlaw}

It is interesting that while the blue-UV slope seems to prefer high
extinctions $E(B-V)\sim 1$, the Balmer decrements seem to prefer
relatively lower values $E(B-V)\sim 0.6$. The extinction law that best
fits the data is the extragalactic law derived from observations of
starburst galaxies \cite{Calzetti2000}. It is commonly believed that
the dust grains in starburst galaxies undergo grain-grain shattering
that results in fragmentation of grains into smaller sub-grains
\citep{Jones1996}. This process effectively results in a size
distribution that is slightly steeper than the standard Mathis, Rumpl,
\& Nordsieck (MRN) size distribution \citep{Mathis1977}. However, the
extragalactic extinction-law is similar to the extinction curve
observed for the small magellanic cloud (SMC) in that the starburst
activity seems to have modified the dust properties in such a way that
has wiped off the small grains responsible for the 2175~$\AA$ bump
\citep[][ and references therein]{Gordon1997}. At the same time, the
small grains resulting in the far-UV extinction rise are not as
depleted as in the SMC case. For Phoenix and other cool-core clusters
with an abundance of X-rays, there is an additional mechanism, namely,
thermal sputtering that disrupts small grains
\citep{Draine1979}. Hence, thermal sputtering could potentially act to
counter the effects of grain-grain shattering, resulting in an
extinction law that is specific to cool-core clusters only. One of our
future goals is to investigate whether any of the existing laws
actually describes cool-core BCGs well enough. We will do so by
allowing the parameters of the extinction law to vary, and including
them in the set of discretely-sampled parameters, $\theta$.

\section{Conclusions}
\label{conclusions}

Previous studies of Phoenix resulted in the highest measurement of
star formation rate for any galaxy, with the estimates reaching up to
$1000~\mpy$. However, the number of models considered in those studies
is very small ($<0.01\%$) relative to those considered in this
work. Estimating star formation rates is a challenging task, where the
main hurdle is exploring the large parameter space. The analysis
essentially entails heavy-duty computation with access to multiple
computing nodes.

Our Bayesian-motivated SED-fitting code, {\bc}, allows us to probe a
large parameter space. Consequently, we are able to explore different
star formation histories, i.e. the constituent stellar populations,
and their physical parameters. This in turns allows us to put
constraints on the values of star formation rates that are
plausible. We consider two models for star formation history,
instantaneous bursts~(SSP) versus continuous star formation~(CSF). In
addition to that we consider a wide range of ages for the old and the
young stellar population with considerable overlap between them. We
find that in the absence of any prior except that the
SFR$< 3000$~$\mpy$ (the predicted cooling-flow rate over the physical
scale of the BCG), the 98\% plausible range for the star formation
rate lies in the range $2230-2890~\mpy$. If we impose a prior on the
extinction, $E(B-V)<0.6$, based on the observational constraints
derived by \cite{McDonald2014} using the Balmer line ratios, the
best-fit SFR value is much lower and in the range $(454-494)~\mpy$. We
regard this as the most probable range of SFR values for Phoenix.

\section*{Acknowledgments} We thank the anonymous referee for their
valuable feedback, time and patience. We thank Prateek Sharma for
useful discussions. R.~M. sincerely thanks the system administrators,
Carsten Aulbert and Henning Fehrmann, at the Albert Einstein Institute
for their generous help in setting up the {\bc} code on the computing
cluster, ATLAS. R.~M. thanks the Albert Einstein Institute for
providing academic support for this work. GB acknowledges support for
this work from the National Autonomous University of M\'exico (UNAM),
through grant PAPIIT~IG100115. M. M. acknowledges support by NASA
through contract HST-GO-13456.002A.

\bibliographystyle{mnras}
\bibliography{ref}

\end{document}